\documentclass{aa}  

\usepackage{graphicx}
\usepackage{txfonts}
\usepackage{xcolor}
\usepackage[breaklinks,colorlinks,citecolor=blue]{hyperref}
\usepackage{soul}
\usepackage{subcaption,caption}
\usepackage{bm}
\usepackage{braket}
\usepackage{hyperref}
\usepackage{adjustbox}
\usepackage{natbib}
\bibpunct{(}{)}{;}{a}{}{,}

\newcommand\be{\begin{equation}}
\newcommand\ee{\end{equation}}
\newcommand\bea{\begin{eqnarray}}
\newcommand\eea{\end{eqnarray}}
\def\ba#1\ea{\begin{align}#1\end{align}}

\newcommand\mr{\mathrm}
\newcommand\bs{\boldsymbol}
\newcommand\mbf{\mathbf}

\newcommand\Omegac{\Omega_\mathrm{cdm}}
\newcommand\Omegam{\Omega_\mathrm{m}}
\newcommand\Omegab{\Omega_\mathrm{b}}
\newcommand\omegac{\omega_\mathrm{cdm}}
\newcommand\omegab{\omega_\mathrm{b}}

\newcommand\invMpc{h/\mr{Mpc}}
\newcommand\Mpc{\mr{Mpc}/h}
\newcommand\Mpcc{[\mr{Mpc}/h]^3}

\newcommand\kmax{k_\mr{max}}
\newcommand\kmin{k_\mr{min}}

\newcommand\kf{k_\mr{F}}

\newcommand\dif{{\rm d}}
\newcommand\cg{\mbf{C}^\mr{G}}
\newcommand\cng{\mbf{C}^\mr{NG}}
\newcommand{\covmos}{\texttt{COVMOS}}
\newcommand{\demcov}{\texttt{DEMNUni-Cov}}
\newcommand{\nercome}{\texttt{NERCOME}}
\newcommand{\halofit}{\texttt{Halofit}}

\newcommand\class{\texttt{CLASS}}

\newcommand\nbk{\texttt{Nbodykit}}

\newcommand\covhf{\texttt{COVMOS\_halofit}}
\newcommand\covdem{\texttt{COVMOS\_demnuni}}

\makeatletter
\newcommand\footnoteref[1]{\protected@xdef\@thefnmark{\ref{#1}}\@footnotemark}
\makeatother

\newcommand{\stef}[1]{\textcolor{purple}{#1}}
\newcommand{\sest}[2]{\setstcolor{purple}\st{#1}\setstcolor{black}\textcolor{purple}{#2}}

\begin{document}

   \title{Cosmological inference including massive neutrinos from the matter power spectrum:  biases induced by uncertainties in the covariance matrix}
   \titlerunning{Covariance}

\author{S. Gouyou Beauchamps\inst{1,2,3}
\and
P. Baratta\inst{1} 
\and
S. Escoffier\inst{1}
\and
W. Gillard\inst{1}
\and 
J. Bel\inst{4}
\and 
J. Bautista\inst{1}
\and 
C. Carbone\inst{5}
}

\institute{Aix Marseille Univ, CNRS/IN2P3, CPPM, Marseille, France
\and Institute of Space Sciences (ICE, CSIC), Campus UAB, Carrer de Can Magrans, s/n, 08193 Barcelona, Spain
\and Institut d’Estudis Espacials de Catalunya (IEEC), Carrer Gran Capitá 2-4, 08034 Barcelona, Spain
\and Aix Marseille Univ, Universit\'e de Toulon, CNRS, CPT, Marseille, France
\and Istituto di Astrofisica Spaziale e Fisica cosmica Milano, Via A. Corti 12, I-20133 Milano, Italy}

   \date{}

% \abstract{}{}{}{}{} 
% 5 {} token are mandatory
 
  \abstract
  {Data analysis from upcoming large galaxy redshift surveys, such as \textit{Euclid} and DESI.
  %Upcoming surveys such as \textit{Euclid} and LSST 
  will significantly improve constraints on cosmological parameters. To optimally extract the maximum information from these galaxy surveys, it is important to control with a high level of confidence the uncertainty and bias arising from the estimation of the covariance that affects the inference of cosmological parameters. In this work, we are addressing two different but closely related issues: 
  %In this paper   we study how accurate and precise the covariance matrix of 2-point statistics needs to be in order to get precise and unbiased cosmological constraints. In this context, two different, but closely related, issues are focused on: 
  (i) the sampling noise present in a covariance matrix estimated from a finite set of simulations and (ii) the impact on cosmological constraints of the non-Gaussian contribution to the covariance matrix of the power spectrum. We focus on the parameter estimation obtained from fitting the full-shape of the matter power spectrum in real space, using 
  %state-of-the-art $N$-body simulations including massive neutrinos. 
  the Dark Energy and Massive Neutrino Universe (\texttt{DEMNUni}) $N$-body simulations. Parameter inference is done through Monte Carlo Markov Chains.
  %, thus allowing posteriors to be non-Gaussian
%(in opposition with usual Fisher matrix analysis). 
 % compared to a Fisher matrix analysis.
  %First, we study the sampling noise in the precision matrix which propagates to parameter space, biasing both the best-fit values and errors of cosmological parameters. We consider two methods to deal with this well known problem and compare their results. The first method, \nercome, an alternative estimator of the covariance matrix based on non-linear shrinkage, is able to significantly reduce the noise induced on the posterior distribution of parameters at the cost of a systematic overestimation of cosmological error bars. The second method, \covmos, aiming the fast generation of approximated catalogues, is able to completely suppress sampling noise effects thanks to a high number of realisations ($10~000$). We show that using a \covmos\ covariance matrix results in unbiased cosmological constraints.
  Regarding the first issue, we adopt two different approaches to reduce the sampling noise in the precision matrix that propagates in the parameter space: on the one hand using an alternative estimator of the covariance matrix based on a non-linear shrinkage, \nercome; and on the other hand employing a method of fast generation of approximate mock catalogs, \covmos. We find that {\nercome} can significantly reduce the noise induced on the posterior distribution of parameters, but at the cost of a systematic overestimation of the error bars on the cosmological parameters. We show that using a \covmos\ covariance matrix estimated from a large number of realisations (10~000) results in unbiased cosmological constraints.
  Regarding the second issue, we quantify the impact on cosmological constraints of the non-Gaussian part of the power spectrum covariance purely coming from non-linear clustering.
  %,often called the non-Gaussian contribution.
  We find that when this term is neglected, both the errors and central values of the estimated parameters are affected for a scale cut $\kmax > 0.2\ \invMpc$. 
  
  %The error on the Hubble parameter $h$ and the neutrino mass $m_\nu$ are underestimated by 20\% and 30\%, respectively, and their best-fit values can undergo a stochastic shift of the order of their error-bars.
  %\stef{est-ce que c'est possbile d'ajouter le mot "Euclid" dans l'abstract? C'est pour le CNES.}\syl{C'est obligatoire ? Je vois pas trop ou caler Euclid dans l'abstract sans que ça fasse name-droping. Euclid est dans l'intro.}

  %Thanks to the very large set of simulations offered by \covmos, we are able to check the validity of previous predictions on sampling noise effects at the level of parameters posteriors. In particular we find that 

  }


   \maketitle
%
%-------------------------------------------------------------------

\section{Introduction}

The main goal of stage-IV 
%galaxy 
dark energy surveys, that are the \textit{Euclid} space mission \citep{Euclid-redbook} and ground-based telescopes like the Dark Energy Spectroscopic Instrument \citep[DESI,][]{aghamousa_16} and the Legacy Survey of Space and Time \citep[LSST,][]{abell_09} at the Vera C. Rubin Observatory,
%like the Legacy Survey of Space and Time \citep[LSST,][]{abell_09} at the Vera C. Rubin Observatory, \textit{Euclid} \citep{Euclid-redbook}, and the Dark Energy Spectroscopic Instrument \citep[DESI,][]{aghamousa_16} 
is to understand the origin of the recent acceleration of the expansion of the universe. 
%For this we need a precise and accurate estimation of cosmological parameters such as the dark energy equation of state parameters or the neutrino mass, to disentangle the various models trying to explain this physical mechanism.
To this end, cosmological parameters such as the equation of state of dark energy, the growth rate of structures or the total neutrino mass are required to be measured at a subpercent level.

%\syl{To understand the origin of the recent acceleration of the expansion of the universe, we need precise and accurate estimations of cosmological parameters such as the dark energy equation of state parameters, the neutrino mass or today's expansion rate, to disentangle the various models that provide different explanations for this daunting problem. For this we will need a large amount of data of great quality that will be provided by ongoing and upcoming dedicated galaxy surveys, like the Legacy Survey of Space and Time \citep[LSST,][]{abell_09} at the Vera C. Rubin Observatory, \textit{Euclid} \citep{Euclid-redbook}, and the Dark Energy Spectroscopic Instrument \citep[DESI,][]{aghamousa_16}.}

To achieve this goal, it is crucial to have a perfect understanding of the covariance of our data-vector as it plays a fundamental role in the statistical inference of parameters. In cosmological surveys, we have only one realisation of the observed data, so the covariance matrix has to be 
%precisely and accurately predicted. 
predicted with precision ("low uncertainty") and accuracy ("low bias").
%Meaning that \wgst{is}{it} 
This means that the covariance matrix has to be noiseless and that its overall amplitude and structure has to be correct, otherwise the estimates of cosmological parameters can be significantly biased in both their values and errors. In practice this can result in wrongly accepting or rejecting a given cosmological model on the basis of the parameter space allowed by this model. In this paper we 
%study how both 
investigate the impact of the precision and the accuracy of the covariance matrix 
%impact the estimation of cosmological parameters,
on the cosmological parameter constraints, considering 2-point statistics of the Large Scale Structure (LSS) in Fourier space, namely the power spectrum $P(k)$. 
%While 
Although these two aspects of the covariance matrix are intrinsically related, we will treat them one after the other in a common context to understand and independently quantify their impact on parameter inference.

%In addition, this parameter is generally unconstrained by present cosmological data so that its posterior distribution 

%This kind of study of the covariance 
Previous works on the impact of covariance uncertainties on cosmological parameters were generally conducted in 
%a standard $\Lambda$CDM context 
the context of the standard $\Lambda$CDM scenario
\citep{sellentin_16, joachimi_16, blot_18,  philcox_20, mohammad_21}, 
%here we want to extend it to
while here we focus on the estimation of the total neutrino mass 
%as 
since the measurement of this fundamental physical parameter is one of the main goals of stage-IV 
%galaxy 
dark energy surveys.
Since the effect of massive neutrinos on the cold dark matter (CDM) density field is only significant on scales smaller than their free streaming length \citep[see][for a review]{lesgourgues_06}, it is mandatory to include scales beyond this limit in the inference of cosmological parameters to achieve a precise measurement of the total neutrino mass. Such small scales are subject to non-linear clustering which complicates the prediction of the covariance matrix of 2-point statistics. Indeed, due to non-linear clustering, the probability distribution function (PDF) of the matter density field becomes non-Gaussian so that higher-order statistics become non-negligible and then contribute to the covariance matrix of the 2-point statistics \citep{bernardeau_01}. 

%Among the difficulties that such evolution of the density field is bringing,

% , in which case we could wrongly rule-out a given cosmological model that does not co

It has been shown that neglecting the non-Gaussian part of the covariance can lead to a significant underestimation of the error bars that we get on cosmological parameters  \citep{barreira_18b, upham_20, lacasa_20}. However, as these non-Gaussian terms involve the 4-point correlation function (or trispectrum in Fourier space) \citep{scoccimarro_99}, their analytical prediction usually requires the use of perturbative methods \citep{wadekar_19} which present some caveats, like the fact that it breaks down on deeply non-linear scales and that it relies on assumptions which are not suited to any cosmological models. An alternative is then to resort to numerical $N$-body simulations which reliably reproduce non-linear clustering. Using numerous realisations of them, one can obtain to some extent a good estimate of the covariance matrix. In addition, this allows to easily estimate the cross-covariance between different probes \citep{bayer_21, taylor_22}, which is not generally guaranteed in the case of current analytic predictions.

%Since $2$-point statistics estimators \phil{do you mean 'estimates' or 'measurements'? I think estimator is reserved to the technique to obtain them...} are usually assumed to follow a multi-variate Gaussian distribution, cosmological parameter inference is generally based on a simple Gaussian likelihood function \phil{I don't see why you are mentioning Gaussian Likelihood, and the connection with the next sentence}. The knowledge of the covariance matrix is then of paramount importance and.

Although $N$-Body simulations are the most accurate way to reproduce non-linear clustering and therefore the non-Gaussian distribution of the density field, they are extremely CPU-time consuming. In consequence it is difficult to create large sets ($\gtrsim 1000$) of high resolution simulations, especially if we want to have a covariance matrix for each cosmological models that are tested. The issue is that with a low sample of mock catalogues, the resulting estimate of the covariance matrix is affected by sampling noise and so is its inverse, the precision matrix, which directly enters the likelihood. This noise is then propagated to the posterior distribution of cosmological parameters, thus generating additional uncertainties. These effects have been thoroughly studied in the literature and theoretical predictions of the amplitude of the additional uncertainties they bring have been provided \citep{taylor_13, dodelson_13, percival_13, taylor_14}. It depends on the number of realisations $N_m$, used to estimate the covariance, but also on the number of data points $N_b$ in the data-vector and the number of free parameters $N_p$. However they all assume that the posterior distribution of parameters is Gaussian, which is generally not the case when one tries to measure the total neutrino mass. Indeed, as the neutrino mass is not well constrained by current cosmological data, its posterior distribution is always cut by the physical prior of a positive mass. 

As the analyses of upcoming galaxy surveys will involve the combination of various cosmological probes (galaxy clustering, weak lensing, CMB-lensing, cluster count...), $N_b$ and $N_p$ will keep increasing, and sampling noise effects will become critical if we want unbiased estimates of cosmological parameters at the percent level \citep{sellentin_16}. Suppressing or drastically reducing sampling noise effects while keeping a good estimate of the non-Gaussian contributions is then one of the great challenges that the analysis of the LSS is facing.

%To get both precise and accurate cosmological inference, 
%where the aim of these surveys is to reduce the error bars on the cosmological parameters, it is essential to find solutions to suppress or reduce drastically sampling noise effects\phil{, while keeping a good estimate of the NG contributions}. 

In the last decade, a vast amount of literature has been devoted to develop solutions to the problem of covariance estimation such as internal estimators\footnote{These are methods designed to estimate the covariance matrix directly from the data.} like 
%jack-knife 
jackknife or bootstrap \citep{escoffier_16, friedrich_15, mohammad_21}, compression methods to reduce $N_b$ instead of increasing $N_m$ \citep{heavens_17, philcox_20}, alternative covariance estimator allowing a reduction of the noise in the precision matrix \citep{pope_07, paz_15}, models with free parameters fitted to simulations \citep{fumagalli_22} or fast 
%approximated 
approximate mock generation methods enabling the creation of very large sets of mock data in a short time \citep{monaco_01, kitaura_13, avila_15, izard_15, agrawal_17}. These methods need to be tested in a realistic parameter inference context to quantify their capability to produce a covariance matrix accurately accounting for the non-Gaussian terms and allowing precise cosmological constraints. This has only been done for the last class of solutions
%(approximated 
(approximate mock catalogues) in \citet{blot_18}.

%The goal of this paper 
%is two-fold. First, we study sampling noise effects and two methods to reduce them: 
%\stef{is to investigate how the estimation of the covariance matrix introduces an additional uncertainty and bias on the estimation of the covariance error in the non-Gaussian case. To this end, we first study the sampling noise inherent to covariance matrix estimation and compare two methods for reducing it:}
%the non-linear shrinkage estimator \nercome\ \citep{joachimi_16} and the fast approximate catalogue generator \covmos\ \citep{baratta_19, baratta_22}. 
%Second, we quantify the impact of the non-Gaussian covariance purely coming 
%\stef{In a second step, we quantify the impact of non-Gaussian covariance}
%from non-linear clustering (referred to as the non-Gaussian term $\cng$ in the rest of the paper) on cosmological parameter 
%estimation. 
%\stef{inference.}

The goal of this paper is two-fold. First, we study sampling noise effects and two methods to reduce them:  the non-linear shrinkage estimator \nercome\ \citep{joachimi_16} and the fast approximate catalogue generator \covmos\ \citep{baratta_19, baratta_22}. Second, we quantify the impact of the non-Gaussian covariance purely coming from non-linear clustering (referred to as the non-Gaussian term $\cng$ in the rest of the paper) on cosmological parameter inference.
Note that we do not consider Super Sample Covariance \citep[SSC,][]{hu_kravtsov_03, takada_hu_13} effects in this work. It has been shown that this term has a significant contribution to weak lensing 2-point statistics covariance \citep{barreira_18b, upham_20, gouyou_21}, but it is not clear whether it is important for spectroscopic galaxy clustering \citep{li_18, li_19, wadekar_20}, {\it i.e.} the 3D power spectrum which we are considering in this work. We leave the inclusion of this other non-Gaussian term in the analysis for future work.

In order to truly understand the impact of the covariance estimation and modeling choices on parameter inference, we need to allow for non-Gaussianity in the posterior of parameters and potential shifts in the best-fit values. Thus, we decide to perform this work using Monte Carlo Markov Chains (MCMC), instead of the Fisher matrix framework which is frequently chosen for this kind of study \citep{joachimi_16, lacasa_19, gouyou_21}. 

Note also that we decide to work with the matter power spectrum in real space, which is far from a real data analysis with galaxy clustering, because we are then lacking redshift space distorsions (RSD) and galaxy bias. However this choice has several advantages. First, we then have a small parameter space to explore as we will only vary 
%4 
four cosmological parameters\footnote{These are the reduced baryon and CDM densities $\omegab$ and $\omegac$ (where $\omega_i \equiv \Omega_i h^2$), the Hubble constant $h \equiv H_{0}/100$ and the neutrino mass $m_\nu$. More detail on the parameter space is given in section \ref{sect:methodo}.} and no nuisance parameters. In this way the MCMC's do not take much computational time so that we can run a large number of them to get a satisfying statistical significance on the effects we are studying. Second, the low dimensionality of the parameter space reduces the potential degeneracies between parameters so that the interpretation of results are easier. Third, we are not subject to biases in cosmological constraints coming from RSD or galaxy bias modeling which is not the focus of this paper. For what concerns sampling noise effects the conclusions we will present in the following (c.f. section~\ref{sect:samnoise} ) should not change much as this noise does not depend on modeling choices or even on the probe under consideration\footnote{Of course, if one considers galaxies in redshift space one must add nuisance parameters and power spectrum multipoles in the likelihood, thus increasing $N_p$ and $N_b$. But this will only change the overall amplitude of sampling noise effects.}. However for the second part of this work, related to the non-Gaussian part of the covariance (c.f. section~\ref{sect:ng_cov}), its relative impact on cosmological constraints could be affected by the inclusion of RSD, galaxy bias and also survey window function that we do not consider either here. The present work can thus be considered as a first step to understand how things work at the matter level in order to better understand the next steps (i.e. galaxies, redshift space and survey window function), which should definitely be considered in future works.

%on comprend mieux ce qu'il se passe, pas de dégénérescences
%plus rapide pour faire beaucoup de fits
%pour le sampling noise ça devrait pas changer les conclusions (seulement le nonmbre de paramètres et de bins)
%pour la non-gaussian covariance c'est intéressant d'y aller step by step, mais c'est sur qu'il faut étendre au reds-space Pgg(k). Leave that for future work.

%{\bf For the parameter estimation we focus in particular on the estimation of the total neutrino mass as its precise measurement is one of the main goal of future galaxy surveys. As discussed above, this is particularly relevant for what concerns \modif{the prediction of sampling noise effects on cosmological parameters, because the estimated posterior distribution of the neutrino mass is generally non-Gaussian (which breaks the main assumption used to derive these predictions).} To do so we perform Monte-Carlo Markov Chain (MCMC) analyses on the real space matter power spectrum of the \texttt{DEMNUni} (for Dark Energy and Massive Neutrino Universe) simulations \citep{castorina_15, carbone_16} \modif{using different sets of sampled covariances}. }

The layout of the paper is as follows. In section~\ref{sect:sim_cov_meth} we 
%present 
describe the $N$-body simulations we used as a reference and 
%its 
the associated covariance matrix, and present the effects that covariance sampling noise has on 
%parameter inference and present 
the inference of cosmological parameters and the two methods we considered to overcome them, \nercome\ and \covmos. In section~\ref{sect:methodo} we present the methodology we followed for parameter inference. In section~\ref{sect:samnoise} we explicitly show the effects of sampling noise on the estimated cosmological parameters and test the reliability of \nercome\ and \covmos\ to recover unbiased estimate of cosmological parameters. We then quantify how the non-Gaussian contribution to the covariance affects cosmological constraints in section~\ref{sect:ng_cov}. 
%We finally 
Finally we conclude in section~\ref{sect:conclu}.

\section{Simulations, and covariances}\label{sect:sim_cov_meth}
    
     In this section we describe the $N$-body simulations and the associated covariance matrix on which the analysis presented in this article is based in section~\ref{subsect:demcov} and \ref{subsect:demcov_cov}. 
     We then introduce the main issues that arise when using a sample covariance matrix for parameter inference in section~\ref{subsect:samp_noise}. Finally  we describe the two methods used in this work to overcome these issues in section~\ref{subsect:covmos} and \ref{subsect:nercome}.
    
    \subsection{The \demcov\ simulations}\label{subsect:demcov}
    
    We use two sets of $N$-body simulations, called the \demcov. Each set corresponds to a different $\Lambda$CDM cosmology, with and without massive neutrinos, with $N_m = 50$ realisations for each. These simulations are part of the \texttt{DEMNUni} simulations project \citep{carbone_16, Parimbelli2021, Parimbelli2022}. They have been ran with the tree particle mesh-smoothed particle hydrodynamics \texttt{GADGET3} which has been modified following \citet{viel_10}, to account for massive neutrinos. The initial conditions are set at $z = 99$, following a method specific to massive neutrino simulations, described in \citet{zennaro_16}. The \demcov\ simulations are characterised by a box side size of comoving length $L = 1000\ \Mpc$  (they have $8$ times smaller volume than the \texttt{DEMNUni} ones) and contain $1024^3$ Cold Dark Matter (CDM) particles, with additional $1024^3$ neutrino particles for the massive neutrino cosmology. The particles have been evolved down to $z=0$ and 5 snapshots have been taken at precisely 
    $$z=\{ 0,\; 0.48551,\; 1.05352,\; 1.45825,\; 2.05053\},$$
    but for simplicity we will quote these redshifts as $z = 0, 0.5, 1, 1.5$ and $2$.
    
    The cosmology with mass-less neutrinos is labeled $0\nu$ and the one with a total neutrino mass of $M_\nu \equiv \sum m_{\nu} = 0.16\ \mr{eV}$ is labeled $16\nu$. The neutrinos mass eigenstates are considered to be degenerate, so that $M_\nu = 3 m_\nu$. The baseline cosmological parameters were chosen according to the 2013 Planck results \citep{planck_13} with zero spatial curvature
    \begin{equation}
        (\Omegam, \Omegab, h, n_s, A_s) = (0.32, 0.05, 0.67, 0.96, 2.1265\times 10^{-9}).
    \end{equation}
    % \begin{eqnarray}
    %     \Omega_\mathrm{m} & = & 0.32\ , \nonumber \\
    %     \Omega_\mathrm{b} & = & 0.05\ , \nonumber \\
    %     h & = & 0.67\ , \nonumber \\
    %     n_s & = & 0.96\ , \nonumber \\
    %     A_s & = & 2.1265\times 10^{-9}\ . \nonumber
    % \end{eqnarray}
    In the two cosmologies $\Omegab$ and $\Omegam \equiv \Omegab + \Omegac + \Omega_{\nu}$ are kept fixed, while $\Omegac$ and $\Omega_{\nu}$ vary to keep $\Omegam$ fixed, according to $\Omega_{\nu} = M_{\nu}/(93.14\ h^{2})$, so that $\Omegac = 0.27$ and 0.2662 for $0\nu$ and $16\nu$, respectively.
    
    We estimate the \demcov\ power spectra in the periodic boxes with the \nbk \footnote{\href{https://nbodykit.readthedocs.io/en/latest/}{nbodykit.readthedocs.io}} \citep{hand_17} software, in real space and directly on the CDM particle field, even for the cosmology with massive neutrino particles. This means that we estimate the $P_\mathrm{cb}(k)$ and not the $P_\mathrm{m}(k)$, where ${\rm cb}$ stands for CDM+baryon and ${\rm m}$ stands for total matter (CDM+baryons+neutrinos).
        
    Regarding the details of the power spectrum estimation, the particles are interpolated on a $1024^3$ grid with a fourth order mass assignment scheme (Piecewise Cubic Spline, or PCS), so that the $k$ range, going from the fundamental mode $k_\mr{F} = 2\pi/L$ to the Nyquist mode $k_\mr{Ny} = \pi/(L/N_s)$ (with $N_s$ the size of the grid) is $[8.10^{-3},\ 3.2]\ \invMpc$. In addition, we switch on the interlacing, to reduce the aliasing for modes close to $k_\mr{Ny}$ and the contribution of the mass assignment scheme is compensated for \citep{sefusatti_15}.
    \begin{figure}
    	\centering
    	\includegraphics[width=0.49\textwidth]{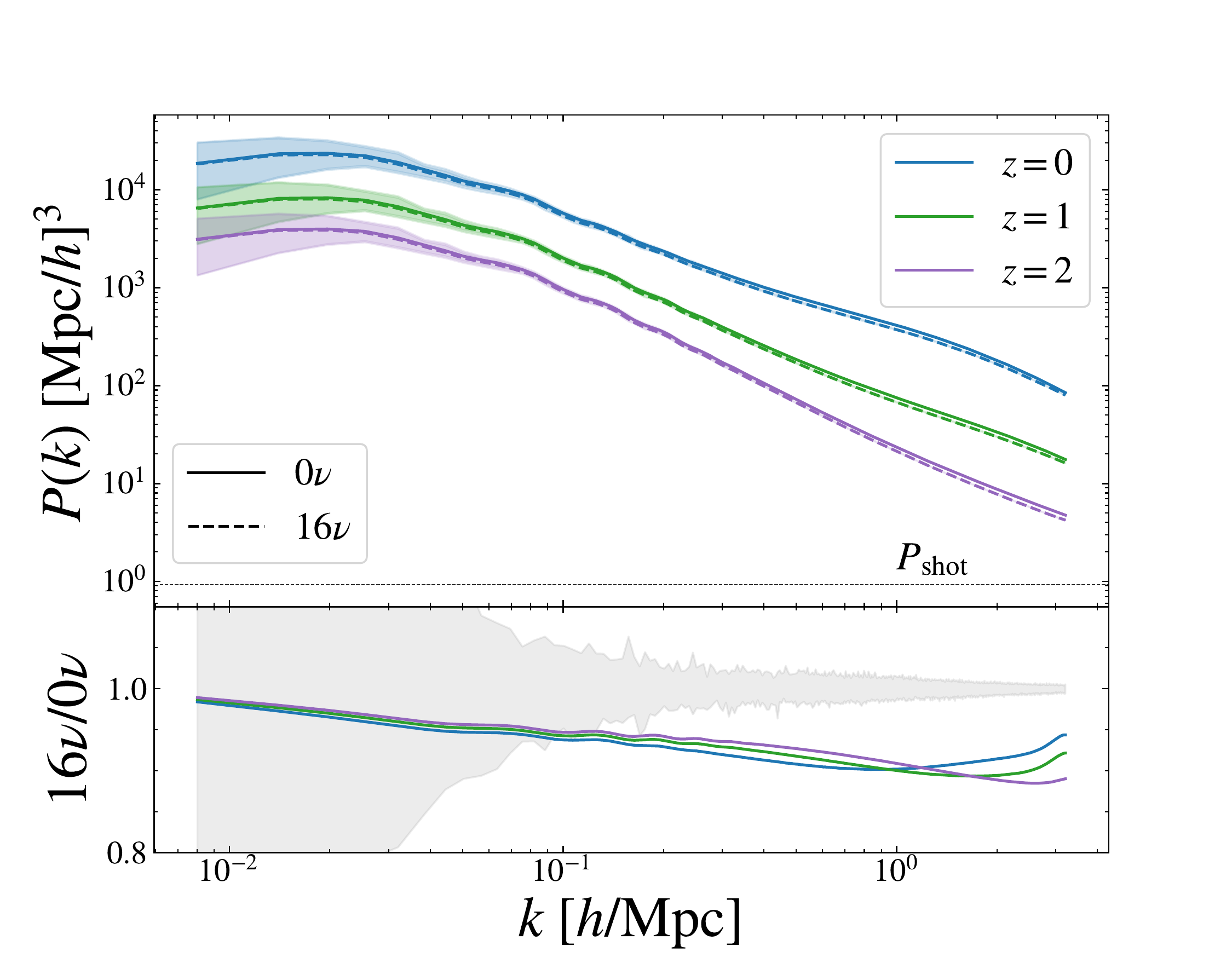}
    	\caption{{\it Top}: The \demcov\ power spectra, for three snapshots and the two cosmologies, $0\nu$ and $16\nu$. The shaded regions show the dispersion over the 50 realisations and the lines are the means on these realisations. The plain and dashed lines corresponds to the $0\nu$ and $16\nu$ cosmology respectively. The black dashed line is the shot-noise level for the $1024^3$ CDM particles. {\it Bottom}: Ratio between the mean power spectrum of the massive neutrino cosmology over the mass-less one. The grey area represents the dispersion over the 50 realisations at $z=0$.}
    	\label{fig:demcov_pk}
    \end{figure}
    
    In 
    %figure~\ref{fig:demcov_pk} 
    Figure~\ref{fig:demcov_pk} we can see the mean and the dispersion over the 50 realisations of the \demcov\ power spectra at $z=0, 1$ and 2 for the two cosmologies. The shot-noise level, referred to as $P_\mathrm{shot} = (L/N_\mr{p})^3 \sim 1\ \Mpcc$, is low compared to the amplitude of the power spectrum, so we neglect it in the following. In the bottom panel, we see the well known effect of massive neutrinos, which damp the amplitude of the power spectrum on small scales. Note that the difference between the two cosmologies becomes significant above $k > 0.1 \invMpc$, where the amplitude of the damping is larger than the dispersion over the 50 realisations.
    
    %The covariance matrix of the power spectrum in a periodic box with a fundamental frequency of $k_\rm{F}$ writes
    %\be
    %\label{eq:pkcov}
    %    C(k_i, k_j) = \dfrac{P^{2}(k_i)}{N_{k_i}}\delta^\mr{K}_{ij} + \kf^3\bar{T}(k_i, k_j) = C^{\mr{G}}(k_i) + C^{\mr{NG}}(k_i, k_j).
    %\ee
    
    \subsection{The \demcov\ covariance matrix}\label{subsect:demcov_cov}
    In the rest  of this article all statistical quantities with a $\hat{ \ }$ will refer to estimated quantities.

    For a given set of realisations of the power spectrum estimated from a sample of independent simulations, an unbiased estimate $\hat{\mbf{C}}$ of the covariance matrix $\mbf{C}$ is
    \begin{equation}\label{eq:cov_estimator}
        \hat{C}_{ij} = \frac{1}{N_m-1}\sum_{n=1}^{N_m} \left[\hat{P}^{(n)}(k_i) - \bar{P}(k_i)\right]\left[\hat{P}^{(n)}(k_j) - \bar{P}(k_j)\right],
    \end{equation}
    where $P^{(n)}$ is the $n$-th realisation of the power spectrum among a total of $N_m$ realisations and where $\bar{P}$ is the mean value of $P^{(n)}$ over the $N_m$ realisations
    \begin{equation}
      \bar{P}(k_i) = \frac{1}{N_m} \sum_{n=1}^{N_m} \hat{P}^{(n)}(k_i).  
    \end{equation}
   Assuming a Gaussian distribution of the power spectrum estimates, the covariance matrix follows a Wishart distribution. Thus, the variance on the covariance matrix elements can be expressed as \citep{wishart_28}
    \be\label{eq:var_cov}
        V[\hat{C}_{ij}] = \dfrac{\hat{C}_{ij}^2 + \hat{C}_{ii}\hat{C}_{jj}}{N_m-1}.
    \ee
    %Using the standard covariance estimator, Eq.~\eqref{eq:cov_estimator}, and with the 50 \demcov\ power spectra of the $16\nu$ cosmology, we estimated the covariance matrix, dubbed $\hat{\mbf{C}}^\mr{D}$ (where the "D" stands for \texttt{DEMNUni}).
    In order to study the impact of non-Gaussianity arising from non-linear processes on parameter estimation, we point out that the covariance matrix is composed of two terms, a Gaussian term $C^{\mr{G}}$  which is diagonal and a non-Gaussian term $C^{\mr{NG}}$ which generates correlations between different Fourier modes \citep{scoccimarro_99}:
    \be
        \label{eq:pkcov}
        C_{ij} = \dfrac{P^{2}(k_i)}{M_{k_i}}\delta^\mr{K}_{ij} + \kf^3\bar{T}(k_i, k_j) = C^{\mr{G}}(k_i) + C^{\mr{NG}}_{ij}.
    \ee
    where $M_{k_i}$ is the number of independent modes in each Fourier shell centered on mode $k_i$, $\delta^\mr{K}_{ij}$ is the Kronecker symbol and $\bar{T}(k_i, k_j)$ is the shell-averaged trispectrum defined by
    \begin{equation}
        \label{eq:trispectrum}
        \bar T(k_i,k_j) \equiv \int \int T(\vec k_1, \vec k_2, -\vec k_1, -\vec k_2) \frac{\dif^3\vec k_1}{V_{k_i}} \frac{\dif^3 \vec k_2}{V_{k_j}},
    \end{equation}
    where the sum is made over two shells centered around $k_i$ and $k_j$\stef{, }  and $V_{k_i}$, $V_{k_j}$ represent their respective Fourier volumes. 
    
\begin{figure}
    	\centering
    	\includegraphics[scale=0.4]{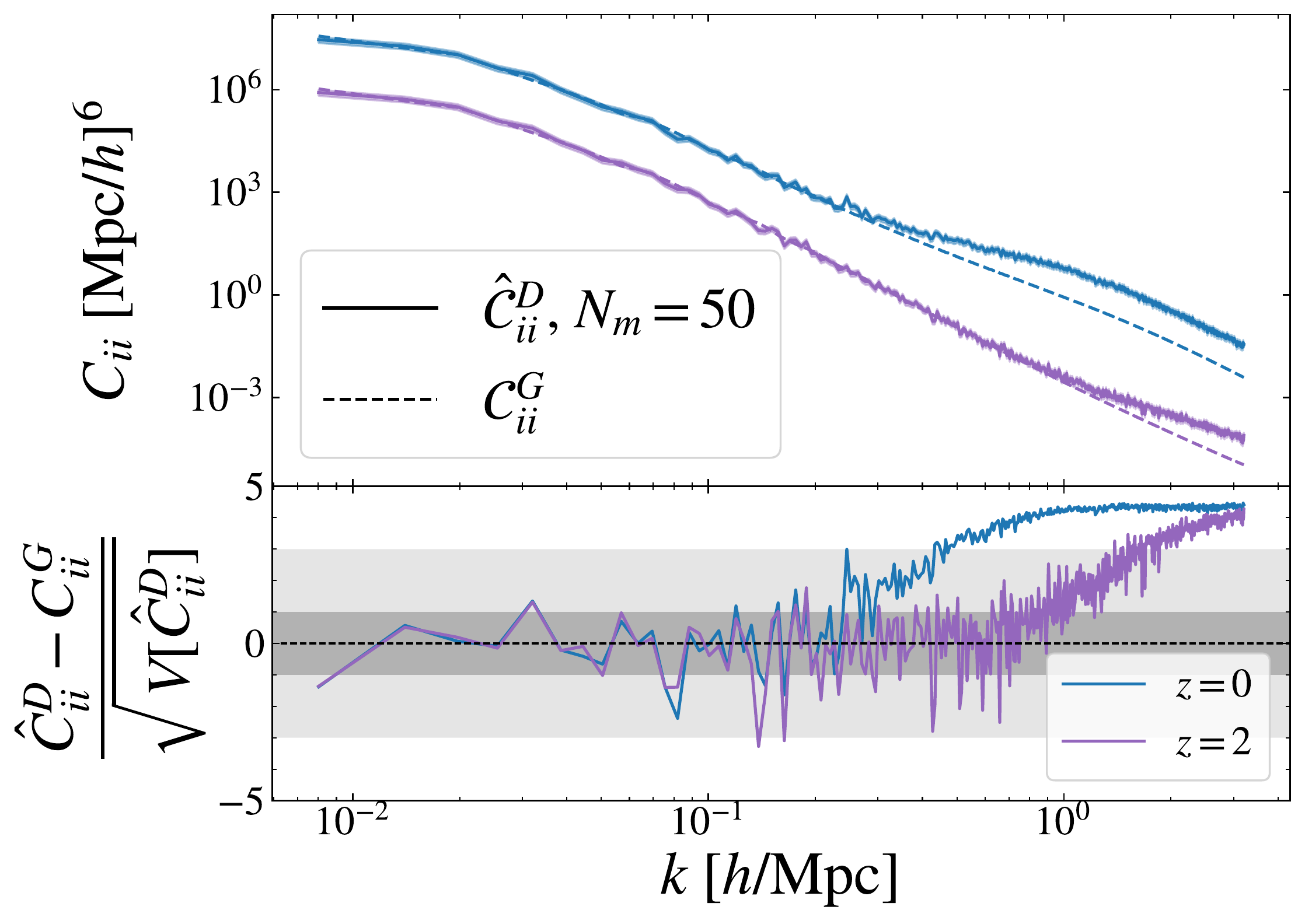}
    	\caption{{\emph{Top}}: Diagonal of the power spectrum covariance matrix in the $16\nu$ cosmology, at $z=0$ (blue) and $z=2$ (purple). Solid lines show the variance estimated from the 50 \demcov\ realisations and the dashed lines display the Gaussian predictions. {\emph{Bottom}}: Residuals between the estimated variance and the corresponding Gaussian prediction normalised by the error on the estimated covariance, determined as in Eq.~\eqref{eq:var_cov}. The grey areas represent the 1$\sigma$ and 3$\sigma$ levels of the residual.}
    	\label{fig:demcov_cii}
    \end{figure}

    Figure~\ref{fig:demcov_cii} shows the diagonal of the estimated covariance matrix using the 50 \demcov\ realisations of the $16\nu$ cosmology, dubbed $\hat{\mbf{C}}^\mr{D}$ (where the "D" stands for \texttt{DEMNUni}), for two redshift snapshots, $z=0$ and $z=2$. For comparison, we show the Gaussian prediction computed from the mean power spectrum, {\it i.e} the $C^{\mr{G}}(k_i)$ term in Eq.~\ref{eq:pkcov}.
    
    We can see the variance excess carried by the trispectrum with respect to the Gaussian approximation, for $k \geq 0.2\ h/\mr{Mpc}$ and $k \geq 0.9\ h/\mr{Mpc}$ at $z=0$ and $z=2$ respectively. This reflects the non-linear evolution of the density field at these scales, which generates non-Gaussian higher-order (larger than 2) correlation functions such as the trispectrum, especially at low redshift. 
    
    %One can see that at $z=0$ at $k\sim 0.3 h/$Mpc the Gaussian covariance is not compatible with the estimated one at the level of $3$-sigma while this is happening only around $k\sim 1h/$Mpc at  $z=2$.
    
    \begin{figure}[ht!]
    	\centering
    	\includegraphics[scale=0.38]{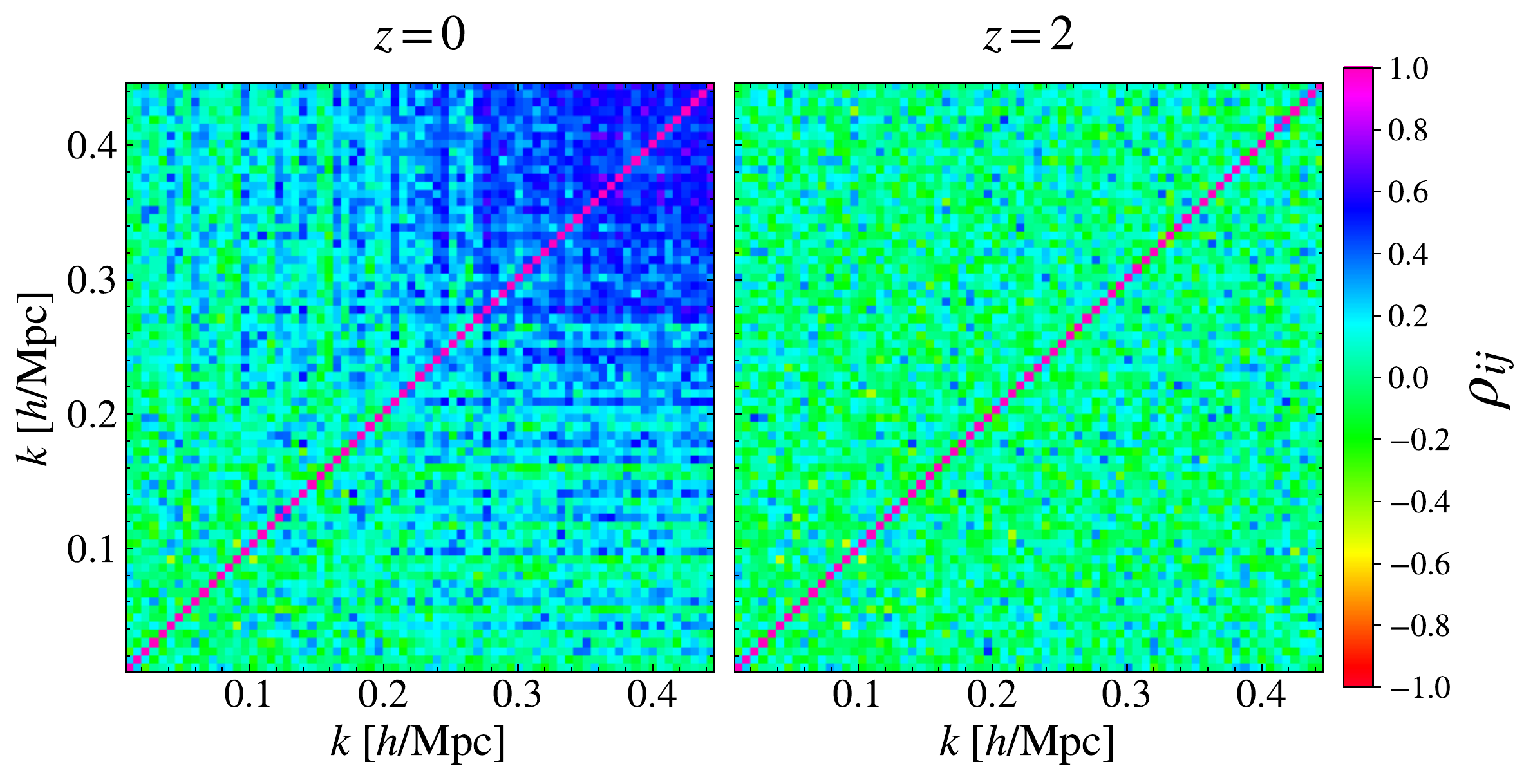}
    	\caption{Estimated power spectrum correlation matrix from \demcov, at $z=0$ (left) and $z=2$ (right).}
    	\label{fig:demcov_cij}
    \end{figure}
    
    Figure~\ref{fig:demcov_cij} exhibits the correlation coefficients $\rho_{ij} = \hat{C}^\mr{D}_{ij}/\sqrt{\hat{C}^\mr{D}_{ii}\hat{C}^\mr{D}_{jj}}$ between the different modes of the power spectrum. Similarly to what we saw on the diagonal elements, the non-Gaussianities ({\it i.e.} the non-diagonal elements) appear at lower $k$ at $z=0$ than at $z=2$. This is due to the non-linear gravitational evolution that propagates up to larger scales for decreasing redshift. More precisely, at $z=2$ and for $k<0.4\ \invMpc$, the correlations are close to zero and consistent with noise, while significant positive correlations are present at $z=0$ for $k>0.2\ \invMpc$.
    
    Furthermore, these two figures show us a large amount of noise in the estimated covariance matrix at all scales due to the low number of available simulations. From Eq.~\eqref{eq:var_cov}, the relative error of the diagonal elements reduces to $\sqrt{V\left[C_{ii}\right]}/C_{ii}$, which only depends on $N_m$ and is equal to $\sqrt{2/(N_m-1)}\sim20\%$ for $N_m = 50$. It is not easy to clearly show the relative error on non-diagonal elements on linear scales, given their relative amplitude with respect to their variance. In other words, the ratio mentioned above blows up for non-diagonal elements because they tend to zero. In this regime, we have seen that the diagonal dominates the covariance matrix, so that noisy non-diagonal elements will not highly impact parameter estimation. However, on mildly non-linear scales ($k>0.2~\invMpc$ at $z=0$), where $C_{ij}^{\rm{NG}}$ starts to be non negligible, the relative noise is close to $50\%$. On these scales, the non-diagonal elements will have an impact on parameter estimation and their noise will then be transferred to parameter posteriors as it will be discussed in the next section and explicitly shown in section~\ref{subsect:param_inf_noise}.
    
    %On the bottom panel we can see the relative error on the covariance matrix. For $k<0.2\ \invMpc$ at $z=0$ and $k<0.4\ \invMpc$ at $z=2$, the relative error can be larger than $100\%$. \ju{What is CijC in the labels ?} This happens because on these scales, the non-diagonal elements tend to zero. In this regime, the diagonal dominates the covariance matrix, so that noisy non-diagonal elements will not highly impact parameter estimation. However, on mildly non-linear scales ($k>0.2\ \invMpc$ at $z=0$), where $C_{ij}$ starts to be non negligible, the relative noise is close to $50\%$. On these scales, the non-diagonal elements will have an impact on parameter estimation and their noise will then be transferred to parameter posteriors as it will be explicitly shown in section~\ref{subsect:param_inf_noise}.
    
    \subsection{Sampling noise in the covariance matrix}\label{subsect:samp_noise}
    
    The statistical quantity involved in parameter estimation is actually the precision matrix $\bs{\Psi} \equiv \mbf{C}^{-1}$. An obvious estimator for $\bs{\Psi}$ is then the inverse of the estimated covariance matrix, $\hat{\bs{\Psi}} = \hat{\mbf{C}}^{-1}$.
    
    As previously stated, the covariance matrix of Gaussian distributed data follows a Wishart distribution, so that its inverse follows an inverse Wishart distribution \citep{wishart_28}. It is then possible to compute the expectation value of the precision matrix estimator, which yields
        \be\label{eq:biased_psi}
            \langle \hat{\bs{\Psi}} \rangle = \dfrac{N_m - 1}{N_m - N_b - 2} \bs{\Psi}, 
        \ee
        where $N_b$ is the number of data points. This is showing that taking the inverse of the estimated covariance matrix in order to estimate the prediction matrix is introducing a predictable bias. It follows that an unbiased estimator of the precision matrix reads
        \be\label{eq:unbiased_psi}
            \hat{\bs{\Psi}}^{U} = \dfrac{N_m - N_b - 2}{N_m - 1} \hat{\mbf{C}}^{-1}.
        \ee
        Since it was first introduced in a cosmological context by \citet{hartlap_06}, this correcting factor is usually called the Hartlap factor.
        
        We see from Eq.~\eqref{eq:biased_psi} that the bias constrains the precision matrix to be estimated with at least $N_m > N_b +2$ and that, even with this condition satisfied, if $N_m$ comparable to $N_b$ this leads to a large bias. Because it is larger than 1, it means that due to a low sample size, the precision matrix is overestimated, {\it i.e.} the covariance is effectively underestimated. The main consequence for parameter inference is then an underestimation of parameter errors. 
        
        Even with an unbiased precision matrix estimator, a small number of realisations leads to a noisy estimate of $\bs{\Psi}$. Let's consider that the estimated precision matrix $\hat{\bs{\Psi}}$ can be decomposed in its true value $\bs{\Psi}$ and the error (or noise) on the truth $\Delta\bs{\Psi}$, such as $\hat{\bs{\Psi}} = \bs{\Psi} + \Delta\bs{\Psi}$. Because it $\Delta\Psi$ originates from a poor sampling of the precision matrix, we call it sampling noise. For Gaussian distributed data, the covariance of $\hat{\bs{\Psi}}$ can be expressed as \citep{taylor_13}
        \be
        \label{eq:psi_cov}
        \langle \Delta\Psi_{ij} \Delta\Psi_{kl} \rangle = A\Psi_{ij}\Psi_{kl} + B(\Psi_{ik}\Psi_{jl} + \Psi_{il}\Psi_{jk}),
        \ee
        with 
        \be
        \label{eq:AB}
            \left\{
            \begin{array}{ll}
                A = \dfrac{2}{(N_m-N_b-1)(N_m-N_b-4)},\vspace{0.3cm}\\ 
                B = \dfrac{N_m-N_b-2}{(N_m-N_b-1)(N_m-N_b-4)}.
            \end{array}
            \right.
        \ee
        
        Using Eq.~\eqref{eq:psi_cov} in the Fisher formalism, several authors have studied how a noisy precision matrix influences the estimation of cosmological parameters, in the case of Gaussian posteriors \citep{taylor_13, dodelson_13, percival_13, taylor_14}. They found that it adds random noise both to the shape and the position of the posterior's maximum.
        
        Let's denote the true error on a parameter $\theta$ as $\sigma_\theta$, the estimation (using a sampled covariance matrix ) of this error as $\hat{\sigma}_\theta$ and the best-fit value as $\hat{\theta}$. The impact of a noisy precision matrix on parameter inference can be summarised in three distinct effects.
        
        (i) There is an additional variance on the variance of the estimated parameters \citep{taylor_13}
        \be\label{eq:taylor_12}
        \mr{Var}[\hat{\sigma}^2_\theta] = \dfrac{2}{N_m - N_b - 4}\sigma_\theta^{4}.
	    \ee
        
        (ii) There is also an additional variance on the position of the best-fit \citep{dodelson_13}
        \be\label{eq:bf_disp}
        \mr{Var}[\hat{\theta}] = B(N_b - N_p)\sigma^2_\theta,
        \ee
        where $N_p$ is the number of parameters to estimate. Note that the direction of this shift in the best-fit is completely stochastic.
	    
	    (iii) The variance of a parameter estimated from the width of the posterior is biased \citep{percival_13}
        \be\label{eq:var_bias}
        \langle \hat{\sigma}^2_\theta \rangle = \left[ 1+A+B(N_p+1) \right] \sigma^2_\theta.
        \ee
        This third effect should not be confused with the first one. The effect (i)  is randomly modifying the size of the errors while (iii) is biasing their average size.

        To account for these last two effects (ii) and (iii), \citet{percival_13} proposed a corrective factor
        \be\label{eq:m1}
            m_1 = \dfrac{1+B(N_b - N_p)}{1+A+B(N_p+1)},
        \ee
        which should be directly multiplied to $\hat\sigma^2_\theta$ when quoting error-bars. In a critical case where $A$ and $B$ are non negligible (i.e. $N_m \sim N_b$), 
        the numerator dominates for $N_b \gg N_p$, while the denominator dominates for $N_b \sim N_p$. In most cases, for cosmological analyses, the number of varied parameters is smaller than the number of data points. It means that the main effect coming from sampling noise is the stochastic shift of the posterior's maximum ({\it i.e.} Eq.~\eqref{eq:bf_disp}).
        
        Presenting the estimated parameter errors, without multiplying by $\sqrt{m_1}$ would not be representative of the actual uncertainty on $\hat{\theta}$, however, as discussed in \citet{wadekar_20}, blindly inflating parameter's error-bars by $\sqrt{m_1}$ has caveats that should be kept in mind when quoting error-bars. First, the resulting corrected covariance matrix of estimated parameters is not representative of the true error on parameters because it corrects for a stochastic shift of the best-fit on average but not for constraints obtained from a single realisation of the precision matrix. Second, if one is interested not only in the confidence regions of the estimated parameters but also in its best-fit values, $m_1$ does not correct for the stochastic dispersion in Eq.~\eqref{eq:bf_disp} as it cannot be predicted. Indeed, it only inflates the error-bars so that, on average, the best-fit lies in it. A last caveat to keep in mind is that the predictions for these additional variance and bias, are only derived for Gaussian distributed parameters and data.
        
        We will compare these predictions against realistic parameter estimations in section~\ref{subsect:nercome}, in particular to test the main assumption needed to derive these expressions which is that the posterior distribution is Gaussian. Even if these predictions are accurate enough so that the additional uncertainties can be rightly accounted for, we certainly want to minimise these uncertainties. In the next two sections we present two methods used in this work aiming the reduction of sampling noise in the precision matrix.

        \subsection{Fast Monte-Carlo catalogues with \covmos}\label{subsect:covmos}
    
%    \sylcol{Summarised description of \covmos\ \citep{baratta_19, baratta_21_in_prep}.}
    
%     \begin{figure*}[t]
%         \centering
%     	\includegraphics[scale=0.4]{figures/covmos_cii_z0.pdf}
%     	\includegraphics[scale=0.4]{figures/covmos_cii_z2.pdf}
%     \caption{Diagonal of the power spectrum covariance matrix at $z=0$ (top) and $z=2$ (bottom) in the $16\nu$ cosmology. \textit{Top}: the black dots represents the \demcov\ covariance, with the associated error-bars in the Gaussian approximation, computed from Eq.~\eqref{eq:var_cov}. The blue and orange lines are the \covmos\ with $N_m=10~000$ and the approximated Gaussian covariance respectively. \textit{Bottom}: Relative difference between the \demcov\ covariance and the two others. The grey shaded area corresponds to the error-bars of the upper panel.}
%     \label{fig:covmos_cii}
% 	\end{figure*}
    The \texttt{COVMOS}\footnote{\href{https://github.com/PhilippeBaratta/COVMOS}{github.com/PhilippeBaratta/COVMOS}} \citep{baratta_19, baratta_22} public code allows the fast realisation of catalogues of various cosmological tracers and models in real and redshift-space. Its fastness stems from the fact that no dynamical evolution is required, as it is usually done in an $N$-body process. Instead, a density field is directly generated at a given redshift, before being submitted to a local Poisson sampling to generate the point-like catalogue.

\begin{table*}
    \centering
        \renewcommand{\arraystretch}{2}
        \caption{ 
        %All the data sets used in this work and their characteristics.
        Characteristics of the data sets used in this work. All sets are available at the 5 redshifts $z= 0, 0.5, 1, 1.5$ and 2.}
        \label{tab:recap}
        \begin{tabular}{| l | p{1.6cm} | p{0.7cm} | p{1.1cm} | p{4cm} | p{6cm} | }
        \hline
        \multicolumn{1}{| l |}{Name} & Cosmology & $n_p$ & $N_m$ & Type & Targeted statistics \\
    	\hline
    	\hline
    	\multicolumn{1}{| l |}{\demcov} & $0\nu$ and $16\nu$ & $1024^3$ & 50 & $N$-body & \\
    	\hline
    	\multicolumn{1}{| l |}{\covhf} & $0\nu$ and $16\nu$ & $10^8$ & 100~000 & Fast Monte-Carlo realisation & \halofit\ $P(k)$ and 1-point PDF estimated from \demcov.\\
    	\hline
    	\multicolumn{1}{| l |}{\covdem} & $16\nu$ & $10^8$ & 10~000 & Fast Monte-Carlo realisation & $P(k)$ and 1-point PDF both estimated from \demcov.\\
    	\hline
        
        \end{tabular}
    \end{table*}

    Although this kind of approach is widely studied in literature through the so-called Log-Normal mock generators \citep[i.e.][]{Xavier:2016elr, agrawal_17}, the originality of \texttt{COVMOS} is that the power spectrum of the density and velocity fields, as well as the density 1-point probability distribution function (PDF), can be arbitrarily set by the user. Thus it is not limited to the Log-Normal form of the PDF and enables more realistic models to be targeted. In particular these statistics can be provided from analytical models, or directly from estimates on simulations or observed data.
    
    Thus specialising in the faithful reproduction of $2$-point statistics, one of the main applications of this method is the production of covariance matrices, in particular those of the power spectrum. It has been shown in \cite{baratta_22} that the produced covariances, when compared with those estimated from $N$-body realisations, are faithfully reproduced in a certain range of scales. In real-space, mode correlations are reliably reproduced up to $k=0.3h/$Mpc for $z\geq 0.5$, while at $z=0$, the certainty interval does not exceed $k\geq 0.17h/$Mpc. Beyond this limit, correlations start to be progressively over-estimated. As some of the subsequent results will exploit smaller scales than this limit at $z=0$, we will keep this information in mind.

    In the present work we created two different \covmos\ data sets:
    \begin{itemize}
        \item The \covhf\ data set which targets the $1$-point density PDF estimated from the \demcov\ simulations and the \halofit\ power spectrum of the \demcov\  cosmology. In this way we perfectly control the output \covmos\ power spectrum. We generated a bit more than 100~000 of these catalogues for the 5 redshifts. We use this data set to study the \nercome\ estimator in section~\ref{subsect:nercome}, sampling noise effect in section~\ref{subsect:param_nercome} and non-Gaussian covariance in section~\ref{sect:ng_cov}.
        
        \item The \covdem\ data set which targets the 1-point density PDF and the power spectrum estimated from the \demcov\ simulations. This set is more realistic in terms of clustering as it reproduces the power spectrum from the $N$-body simulations. We generated 10~000 of these catalogues for the 5 redshifts. The same kind of setting was used in \cite{baratta_22} to validate \covmos\ against the \demcov\ simulations at level of the power spectrum and its covariance matrix. We used this data set to study the accuracy of the \covmos\ covariance matrix at the level of parameter constraints in section~\ref{subsect:param_covmos}.
    \end{itemize}
    Both sets are characterised by negligible shot-noise compared to the measurements ($n_p = 10^8$ particles per realisation). They have been generated on periodic comoving volumes of size $L=1000\ \Mpc$, with the \texttt{COVMOS} sampling parameter $N_s=512$ (corresponding to the spatial resolution of $2\ \Mpc$). The entire sample of simulation and associated power spectra estimation took $\sim 5$ days, running on about $940$ processors of $2.4$ GHz of the \texttt{Dark Energy Centre} in Marseille, France. For a recap, Table~\ref{tab:recap} presents all the data sets that are used throughout the article and their characteristics.

    \subsection{Non-linear shrinkage with \nercome}\label{subsect:nercome}
    
    The \nercome\ (Non-parametric Eigenvalue Regularised Covariance Matrix Estimator) algorithm was first proposed by \citet{lam_16} and then applied in a cosmological context by \citet{joachimi_16}. This estimator is designed to reduce the bias and the variance present in an estimated precision matrix, which propagates to parameter estimation. 
    
    The \nercome\ algorithm can be decomposed in 3 steps. Let's consider a set of $N_m$ realisations of the data vector $\mbf{X}$ of size $N_b$, from which we want to estimate the covariance. 
    \begin{enumerate}
        \item Divide the set of realisations in two subsets of size $s$ and $N_m-s$ respectively.
        \item Apply the standard estimator to each subset to obtain the covariance matrices $\hat{\mbf{S}}_i$, with $i=1$ or 2. Decompose them in the form $\hat{\mbf{S}}_i = \mbf{U}_i\mbf{D}_i\mbf{U}_i^{T}$, with $\mbf{U}$ the matrix of eigenvectors and $\mbf{D}$ the diagonal matrix of eigenvalues.
        \item Estimate the covariance matrix as
            \be\label{eq:nercome}
                \hat{\mbf{C}}_\mr{N} \equiv \mbf{U}_1\mr{diag}(\mbf{U}_1^{T}\hat{\mbf{S}}_2\mbf{U}_1)\mbf{U}_1^{T},
            \ee
        and average $\hat{\mbf{C}}_\mr{N}$ over $N_c$ different random compositions of the subsets and a fixed split-position $s$.
    \end{enumerate}
    In the original version of the algorithm the operation is repeated for different $s$ to find the optimal split, but here, following \citet{joachimi_16}, we will keep a fixed $s = (2/3)N_m$ and $N_c = 500$.

    An insight to understand Eq.~\eqref{eq:nercome}, is that taking the diagonal of $\mbf{U}_1^{T}\hat{\mbf{S}}_2\mbf{U}_1$, which mixes two matrices estimated from independent data sets, will shrink both large and small eigenvalues to avoid singular values\footnote{This shrinkage is said to be non-linear because it shrinks different eigenvalues by different amounts. See \citet{pope_07} for a reference on linear shrinkage.}, so that $\hat{\mbf{C}}_\mr{N}$ is always positive definite.
    
    Following \citet{joachimi_16}, we can test the efficiency of $\nercome$ in reducing the bias and the variance in the estimated precision matrix, by considering the signal-to-noise ratio (S/N), defined as
    \be\label{eq:snr_ner}
        F \equiv (\mr{S/N})^2 = \sum_{i,j} P(k_i)\Psi_{ij}P(k_j).
    \ee
    If it is estimated using a Hartlap-biased precision matrix, then the expectation value yields $\langle \hat{ F} \rangle = \left[(N_m-1)/(N_m-N_b-2)\right] F$. So it is possible to measure the bias in $\hat{\bs{\Psi}}$\sest{,}{} by performing an average of the SNR over a certain number of realisations of the precision matrix\sest{,}{} and to divide by the SNR computed with the true precision matrix, such that 
    \be\label{eq:SNR_hartlap}
        \dfrac{\langle \hat{ F} \rangle}{ F} = \dfrac{N_m-1}{N_m-N_b-2}. 
    \ee
    In the same way, one can show that
    \be\label{eq:SNR_var}
        \dfrac{\sqrt{\langle \hat{ F} - \langle \hat{ F}^2 \rangle \rangle}}{ F} = \dfrac{\sqrt{2}(N_m-1)}{\sqrt{N_m-N_b-4}(N_m-N_b-2)}
    \ee  
    is equivalent to the variance of the estimated precision matrix \citep{taylor_14}. Thanks to the $10~000$ \covdem\ catalogues that we presented in the previous section, we can compute Eq.~\eqref{eq:SNR_hartlap} and \eqref{eq:SNR_var} for several $N_m$ and compare the result between the standard and \nercome\ estimator.
    
    \begin{figure}
    	\centering
    	\includegraphics[scale=0.45]{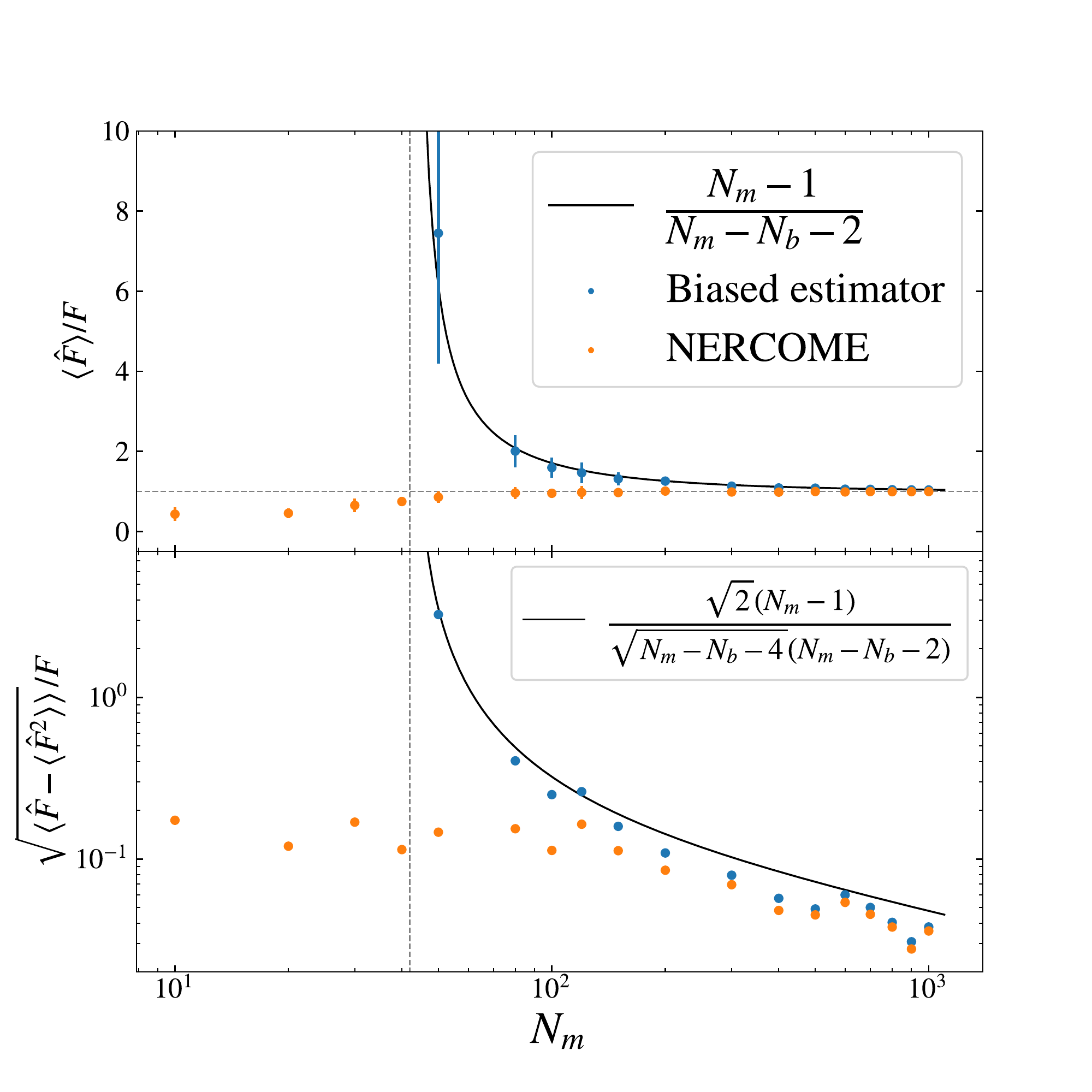}
    	\caption{Bias (top panel, Eq.~\eqref{eq:SNR_hartlap}) and variance (bottom panel, Eq.~\eqref{eq:SNR_var}) in a precision matrix estimated with $N_m$ mocks, for $N_b = 40$. The blue and orange dots respectively correspond to the biased ({\it i.e.} without the Hartlap correction) and \nercome\ estimator. The black line represents the prediction of the two quantity, for Gaussian distributed data. The vertical dashed grey line indicates the Hartlap limit $N_m \leq N_b + 2$.}
    	\label{fig:SNR_nercome}
    \end{figure}
    
    For the data vector in Eq.~\eqref{eq:snr_ner}, we took the mean power spectrum over the $10~000$ realisations and cut it to $N_b=40$. We assumed the precision matrix estimated with the $10~000$ mocks to be the true one and we averaged over 10 realisations of $\hat{\bs{\Psi}}$ for each $N_m$. The result is displayed in 
    %figure~\ref{fig:SNR_nercome}. 
    Figure~\ref{fig:SNR_nercome}. 
    Looking at the bias on the precision matrix (top panel), the prediction is well within the error-bars for the biased estimator ({\it i.e.} without the Hartlap correction), and diverges for $N_m<N_b+2$. We see that for these $N_m$, \nercome\ is able to produce a non-singular covariance matrix, which however shows a slight bias, lower than 1. For larger $N_m$, the bias is compatible with 0. In the case of the variance of the precision matrix (bottom panel), the standard estimator is also well in agreement with the prediction. Like for the bias, \nercome\ largely reduces the variance in the precision matrix, especially at low $N_m$. For $N_m>100$, \nercome\ results in a variance close to the biased estimator, but systematically lower.

    These results are in agreement with \citet{joachimi_16}, which however used a Jack-Knife technique to produce several realisations of the precision matrix. We will test the reliability of a covariance matrix estimated with \nercome\ at the level of parameter estimation in section~\ref{subsect:param_nercome}.

\section{Parameter inference methodology}\label{sect:methodo}
    In this section we present the methodology analysis choices we followed for parameter inference using MCMC.
    
    \subsection{Likelihood}
    
    By making the assumption that the estimator of the power spectrum $\hat P(k)$ follows a multivariate Gaussian distribution, the Likelihood $L(\hat{\bs{P}}(k); \bs{\theta})$ has the following form 
    \be
    \label{eq:lik_dem}
        -2\log{L} = [\hat{\bs{P}}(k) - \bs{P}(k; \bs{\theta})]^T \mbf{C}^{-1}  [\hat{\bs{P}}(k) - \bs{P}(k; \bs{\theta})] \equiv \chi^2,
    \ee
    where $\hat{\bs{P}}(k)$ is the data vector, $\bs{\theta}$ is the set of free cosmological parameters,  $\bs{P}(k; \bs{\theta})$ is the theoretical prediction depending on the cosmological parameters, and $\mbf{C}$ is the covariance matrix. Throughout this article we will use different data vectors and covariance matrices, chosen among the different simulation sets presented in the previous section, according to which effects we want to study. However we always consider the same theoretical prediction.
    
    We use \halofit\ \citep{smith_02} to model the non-linear CDM+baryon power spectrum. The version of \halofit\ we use incorporates both the treatment of massive neutrinos developed by \citet{bird_11} and the revised fitting-formulae by \citet{halofit_takahashi_12}. We use \class\ \citep[][version 2.9]{blas_11} to compute it.
    
    We chose to simultaneously fit the 5 redshifts of \demcov\ to include the information of redshift evolution so that we avoid too large degeneracies between cosmological parameters. To circumvent large correlations between the different redshifts we select realisations with different initial conditions for each redshift\footnote{Indeed, for a given realisation of a simulation box the different redshifts of this box are highly correlated, especially on large scales, as all the snapshots inherit from the same initial conditions.}. As a consequence, the $\chi^{2}(z_{i})$ at the different redshifts are uncorrelated and the full $\chi^2$ can be expressed as
    \be
        \chi^{2} = \sum_{i=0}^{n_{z}=5} \chi^{2}(z_{i}).
    \ee
    Thus when the covariance used for parameter inference is the one from \demcov, it can only be estimated with the 45 remaining realisations. 
    
    The MCMCs are always performed in the range $\kmin = k_\mr{F}=2\pi/L \approx 0.01\ \invMpc$ to $\kmax$ (which will be varied), with a step $\Delta k = k_\mr{F}$. The cosmological parameters that are varied are the reduced baryon and CDM densities $\omegab$ and $\omegac$\footnote{Where $\omega_i \equiv \Omega_i h^2$.}, the Hubble parameter $h \equiv H_{0}/100$ and the single neutrino mass $m_\nu$\footnote{As stated in section~\ref{subsect:demcov} the $16\nu$ cosmology assumes a degenerate mass hierarchy, so that it makes no difference to fit for $M_\nu$ or $m_\nu$.}. The priors on these parameters are chosen to be broad and uniform to avoid further effects on parameter inference output. They are listed in 
    %table~\ref{tab:priors} 
    Table~\ref{tab:priors}
    along with their fiducial values. We also consider the case of a tighter Gaussian prior coming from Big Bang Nucleosynthesis (BBN) for $\omegab$ in section~\ref{sect:ng_cov}. We use \texttt{MontePython} \citep[][version 3.3.0]{montepython_3} to run the MCMC and \texttt{GetDist} \citep{get_dist} to produce the plots of the posterior densities. We assert the convergence of the chains with a Gelman-Rubin criterion \citep{gelman_92} : $R-1 < 0.01$.
    
    \begin{table}
    \centering
     \caption{Uniform priors for cosmological parameters and fiducial values for the $16\nu$ cosmology.}
        %\begin{adjustbox}
        \begin{tabular}{l | c c c }
        \hline
        \multicolumn{1}{l |}{$\bs{\theta}$}  & Priors & Fiducial value\\
    	\hline
    	\hline
    	\multicolumn{1}{l |}{$\omegab$} & $[0.01,\ 0.06]$ & 0.0224\\
    	\multicolumn{1}{l |}{$\omegac$} & $[0.01,\ 0.08]$ & 0.1195\\
    	\multicolumn{1}{l |}{$h$} & $[0.3,\ 1.5]$ & 0.67\\
    	\multicolumn{1}{l |}{$m_{\nu}\ [\mathrm{eV}]$} & $[0,\ 1]$ & 0.0533 \\
    	\hline
        
        \end{tabular}
        %\end{adjustbox}
        \label{tab:priors}
    \end{table}
    
    \subsection{Observable space}\label{subsect:obs_space}
    
    Given current uncertainties on the measurement of $H_0$, cosmological distances are usually expressed in $\mr{Mpc}/h$. This implies that when fitting for the full-shape of the power spectrum with $h$ being one of the free parameters, one has to take care of also varying $h$ in the overall amplitude of the power spectrum and in the $k$ grid on which the theory vector is generated. Indeed, both $P(k)$ and $k$ depend on $h$ through their units, which are respectively $[\mr{Mpc}/h]^3$ and $\invMpc$. This dependence actually expresses what can be referred to as the isotropic Alcock-Paczyński (AP) effect, due to the fact that we need to assume a fiducial cosmology in order to get comoving distances \citep[see][for more discussion on this]{sanchez_20}. %What happens in practice is that, for a given set of cosmological parameter, \class\ (its \texttt{python} wrapper actually) gives 
    
    However, in the case of simulated data we know the value of $h$, so that we can express both our measurement and theory vectors in units of $\mr{Mpc}$ by using this known value. By doing so we remove part of the dependence in $h$ of the power spectrum. In appendix~\ref{app:mpch} we compare the shape of the posterior distribution in both cases and observe that when accounting for the isotropic AP effect the posterior distribution can present non-Gaussian features due to some degeneracies between the parameters brought by the additional $P(k)$ dependence in $h$ through this effect. 
    
    In section~\ref{sect:samnoise} we are interested in the effect of sampling noise on the parameters posterior distribution, thus we need to control the shape of the posterior and in particular control whether it presents or not non-Gaussian features. Hence for that section we chose to perform our fits in the Mpc$^{-1}$ space, {\it i.e.} not accounting for the isotropic AP effect.
    
    Even though this way of performing a $P(k)$ full-shape fit is not representative of what we should do with real data, we recall that this choice does not impact the conclusions we take from the results presented in section~\ref{sect:samnoise}, as they focus on the effect of sampling noise in the covariance. As stated in the introduction, this noise does not depend on modeling choices or even on the probe under consideration. We will come back to this discussion and modify this analysis choice in section~\ref{sect:ng_cov}.

\section{Sampling noise effects on cosmological parameter estimation}
\label{sect:samnoise}
In this section we show how the noise in the covariance matrix influences the shape of the cosmological parameters posterior distribution, and test the reliability of \nercome\ and \covmos\ to overcome these effects.

%After briefly presenting our setting and methodolgy for the MCMC, we will explicitly show how sampling noise affects the best-fit and error of the estimated parameters in the case of a high sampling noise. 

%Then we will test the performance of \nercome\ in reducing the effect of sampling noise and the reliability of the covariance matrix estimated with \covmos\ mocks.
    
    \subsection{Parameter inference with a high sampling noise}\label{subsect:param_inf_noise}
    To explicitly show the impact of sampling noise on parameter inference, we perform a fit of the \demcov\ power spectrum in the 16$\nu$ cosmology at the 5 redshifts using the Gaussian ($\mbf{C}^\mr{G}$) and the \demcov\ ($\hat{\mbf{C}}^\mr{D}$) covariance matrices. As in the Gaussian case we deal with an analytic covariance, the inversion of the matrix is not impacted by the number of modes per redshift, so we chose a range of $\kmax = [0.1,\ 0.275]\ \invMpc$ corresponding to $N_b = [16, 44]$. However, when estimated from simulations, the precision matrix is constrained to be estimated with $N_m > N_b +2$ (c.f. Eq.~\eqref{eq:biased_psi}). To keep the same wave-mode binning, in the \demcov\ case the $\kmax$ range is restrained to $[0.1,\ 0.25]\ \invMpc$, corresponding to $N_b = [16, 39]$. For larger $N_b$, the inversion of the covariance matrix is numerically unstable.
    
    Note that here we don't aim at studying the difference in parameters constraints due to the addition of the non-Gaussian contribution contained in $\hat{\mbf{C}}^\mr{D}$ as this will be discussed in section~\ref{sect:ng_cov}. Here, $\mbf{C}^\mr{G}$ serves as a reference given that it is noise-free. 
    
    \begin{figure}
    \centering
    \includegraphics[scale=0.46]{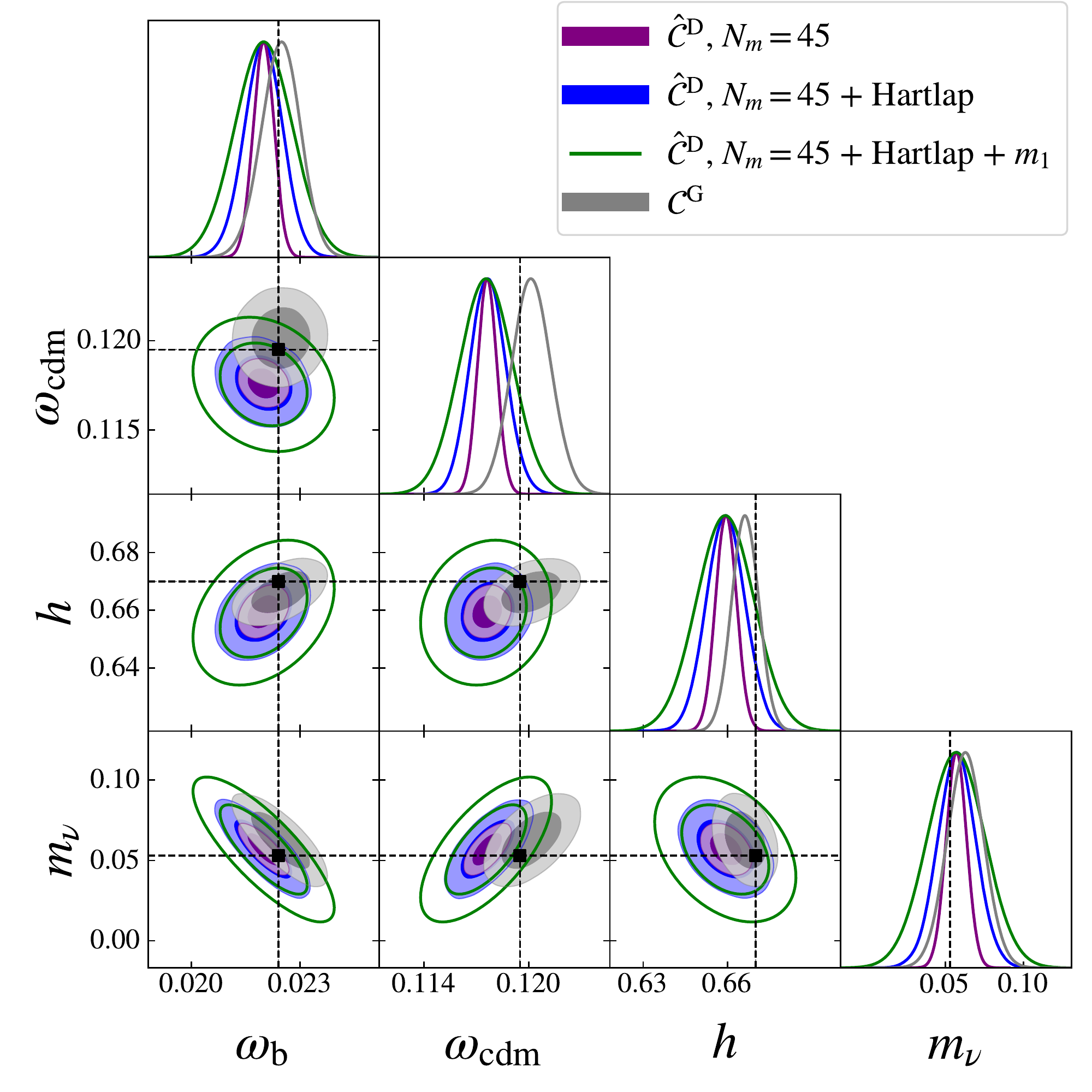}
    \caption{2D and 1D marginalised posteriors, obtained for $\kmax=0.19\ \invMpc$, with the analytic Gaussian covariance (grey), the covariance estimated from $45$ \demcov\ without the Hartlap correction (purple), with the Hartlap correction (blue) and with $m_1$ factor (green). For the 2D posteriors, the 68.3\% and 95.5\% confidence regions are shown. The black square and dashed lines show the fiducial cosmology.}
    \label{fig:triangle_high_samnoise}
    \end{figure}
    
    Figure \ref{fig:triangle_high_samnoise} shows the shape of the 2D posteriors obtained for $\kmax = 0.2\ \invMpc$ with the Gaussian covariance and the \demcov\ covariance, corrected or not for the Hartlap bias, as well as the posterior distribution inflated by the $m_1$ factor. We can see that it is only in the Gaussian covariance case that all parameters are recovered within the 1$\sigma$ limit, in all the other cases the ellipses are shifted away from the fiducial cosmology. As the model is the same in all 4 cases, this shift is only due to the fact that the covariance matrix is estimated with a low $N_m$ in the 3 cases using $\hat{\mbf{C}}^\mr{D}$. Comparing the width of the posteriors we can observe different effects that were explained in section~\ref{subsect:samp_noise}. Without the Hartlap correction on $\hat{\mbf{C}}^\mr{D}$, the posterior width is well underestimated, in comparison with $\mbf{C}^\mr{G}$. When correcting for the Hartlap bias, one can observe that the width of the posterior is increasing to reach a similar size as in the Gaussian case.  Nevertheless, the best-fit doesn't change and is still shifted by more than 1$\sigma$ with respect to the fiducial cosmology for most parameters. By further inflating the errors by $\sqrt{m_1}$, we can gauge the effect of the noise in $\hat{\bs{\Psi}}$, transferred to the cosmological parameters, which is not accounted for by the Hartlap factor. This huge error-bar accounts for the stochastic shift in the position of the best-fit that was discussed above.
    
\begin{figure}
    \centering
    \includegraphics[scale=0.48]{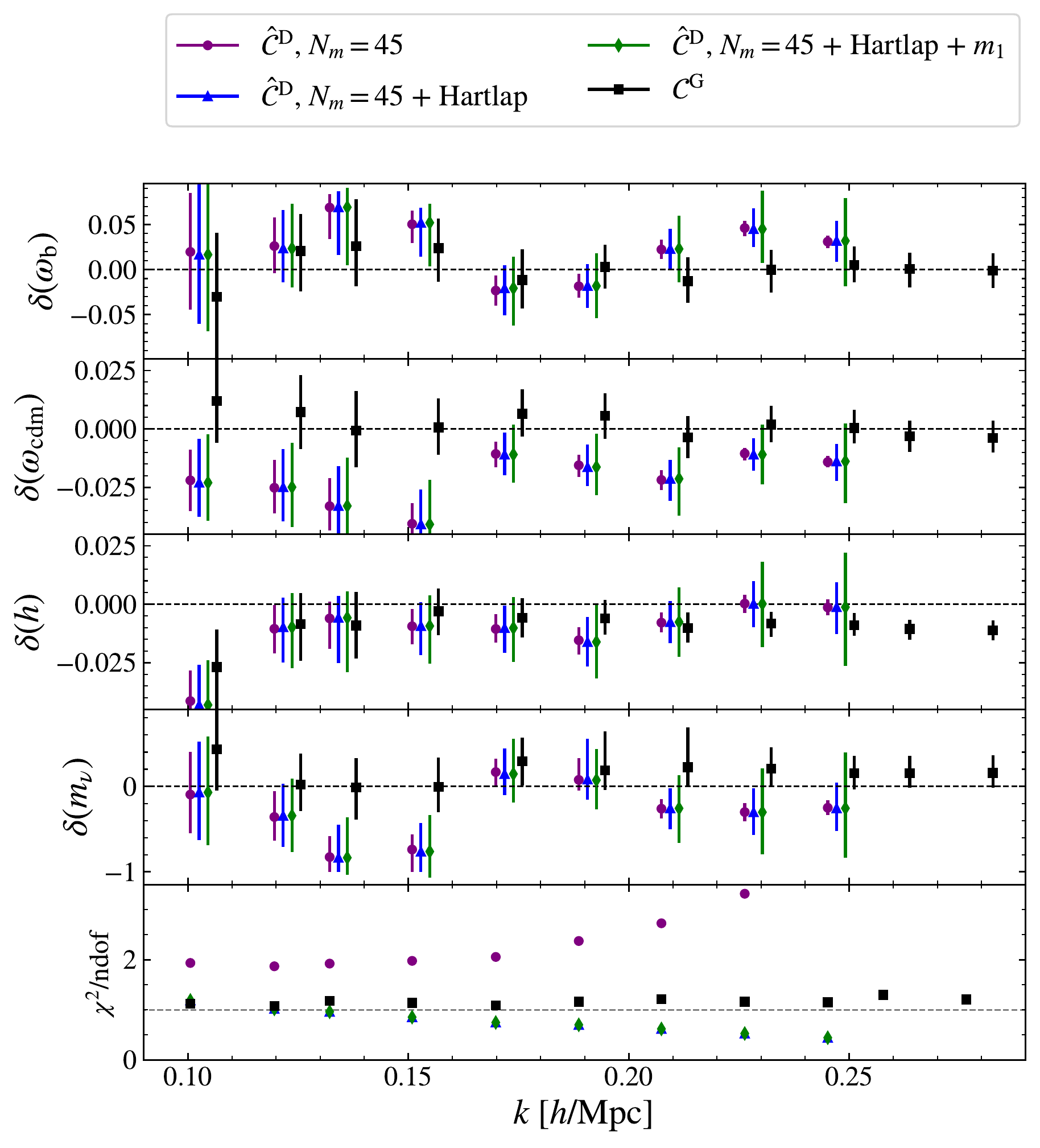}
    \caption{Marginalised parameter constraints, derived from the same cases as in figure~\ref{fig:triangle_high_samnoise}. The four top panels represent the relative difference with the $16\nu$ cosmology for each free parameter. The bottom panel shows the $\chi^2$ over $n_\mr{dof}$. On this last panel the green and blue dots overlap because the difference is just in the presence of $m_1$ which does not affect the $\chi^2$. }
    \label{fig:dev_m1}
    \end{figure}

    To further gauge the effect of sampling noise, we show in Figure~\ref{fig:dev_m1} the marginalised parameter constraints for the 4 cases as well as $\chi^2/n_\mr{dof}$ with respect to $\kmax$, where $n_\mr{dof} = N_b - N_p$ is the number of degrees freedom. We start by commenting the 4 top panels corresponding to the parameter constraint and then we focus on the $\chi^2$ panel.
    
    \begin{figure}[ht]
    \centering
    \includegraphics[scale=0.5]{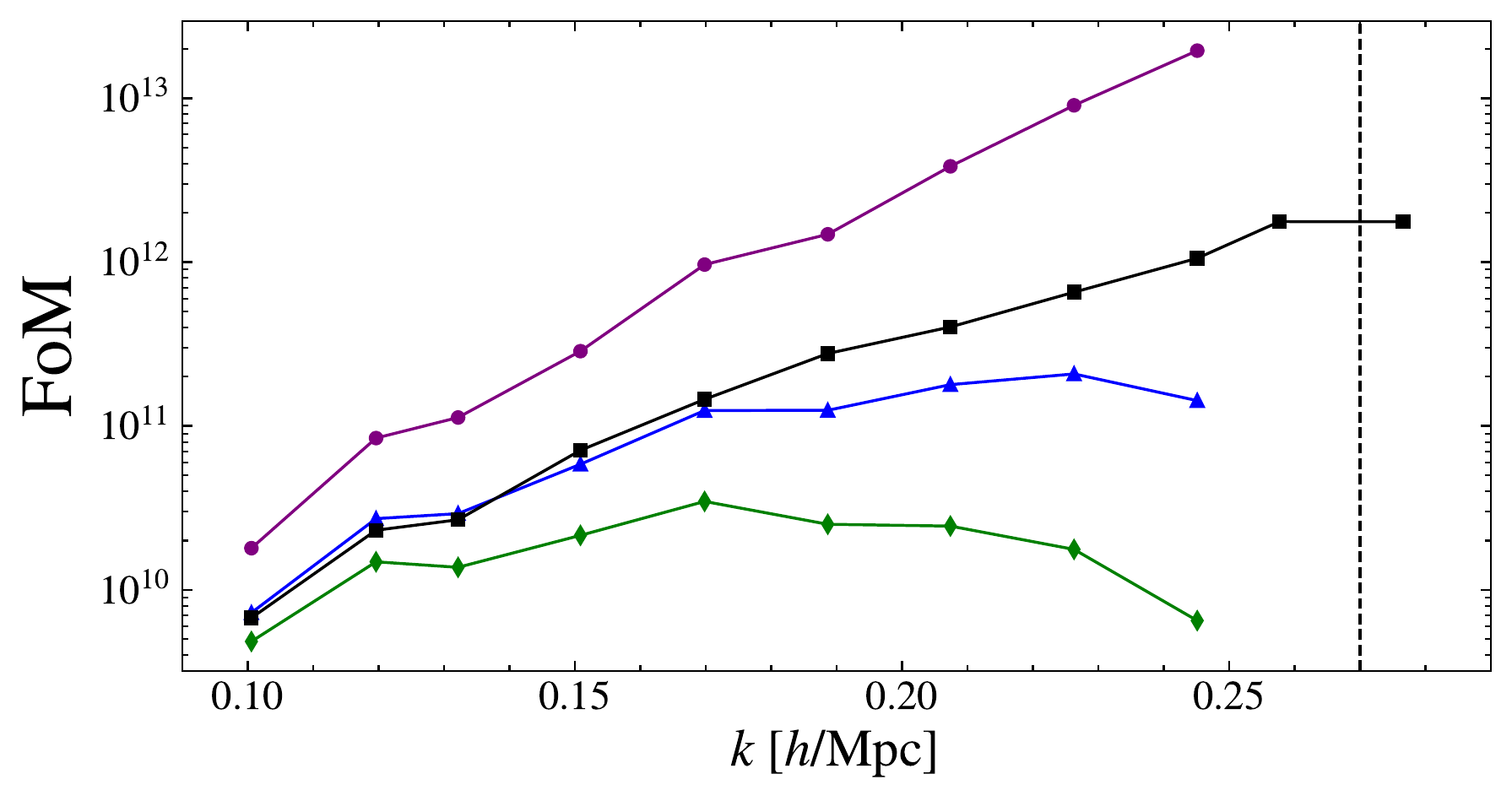}
    \caption{FoM obtained from the parameter constraints in the $16\nu$ cosmology for the same cases as %figure~\ref{fig:triangle_high_samnoise} 
    Figures~\ref{fig:triangle_high_samnoise}
    and \ref{fig:dev_m1}. The 
    %black 
    vertical dashed line shows the $\kmax$ corresponding to the limit $N_b = N_m - 2$, above which the covariance matrix estimated with $N_m = 45$ cannot be inverted.}
    \label{fig:fom_m1}
    \end{figure}
    
    As the Gaussian covariance is noise-free, the fluctuations in the constraints that we see while varying the $\kmax$ are only due to the noise present in the data vector. Hence, it can serve as a reference to observe the noise in the covariance. For $\kmax \sim 0.1\ \invMpc$, the variance of the power spectrum is important, leading to biases close or larger than 1$\sigma$ in the estimated parameters, especially for $h$. The error-bars decrease and the best-fit shifts toward the true cosmology with increasing $\kmax$. But for $\kmax > 0.2\ \invMpc$, \halofit\ starts to fail in reproducing the non-linear power spectrum, leading to a systematic bias in the estimation of $m_\nu$ and $h$.
    
    Looking at the constraints obtained with the \demcov\ covariance, we observe larger fluctuations through the $\kmax$ range, compared to the Gaussian case, exhibiting the effect of sampling noise. Looking at the error bars, we can see the same effects that were discussed above with the Hartlap and $m_1$ factor. Here we can appreciate how the intensity of these effects grow with $\kmax$ as $N_b$ increases in the same time.
    
    The $\chi^2/n_\mr{dof}$ is stable and remains close to 1 in the case of the Gaussian covariance, indicating a good-fit. However, it is overestimated by a factor of 2 (or more for large $\kmax$) in the case of the \demcov\ covariance without the Hartlap correction. Conversely, when accounting for the Hartlap bias, the $\chi^2$ is lower than 1. This result was explained by \citet{sellentin_15} who showed that the $\hat{\chi}^2$ estimated with a Hartlap-biased precision matrix is drawn from a distribution with a larger width than a true $\chi^2$ distribution. On the contrary when it is corrected for the Hartlap bias, the distribution is sharper. 
    
    Finally in Figure~\ref{fig:fom_m1},
    we expose the FoM computed from the above constraints, following the same color code as in the two previous figures. Again, we clearly see the overestimation of the FoM when the Hartlap bias is not accounted for. For $\kmax~<~0.15\ \invMpc$, the Hartlap factor seems sufficient to recover the FoM obtained with the Gaussian covariance, which can be considered as a reference at these scales. But, as we can see in Figure~\ref{fig:dev_m1}, the best-fits present large biases on those scales, especially $h$ and $m_\nu$. This is illustrated by the decrease of the FoM caused by the $m_1$ factor, which accounts for this bias. Furthermore, note that with both corrections, the FoM decreases as the $\kmax$ is getting close to the limit $N_k = N_m - 2$. In that case the Hartlap bias and sampling noise are so high that increasing the amount of information in the fit by increasing the $\kmax$ degrades the constraints. This is exactly what we want to avoid.  

    \begin{figure*}[h]
        \centering	
        \includegraphics[scale=0.35]{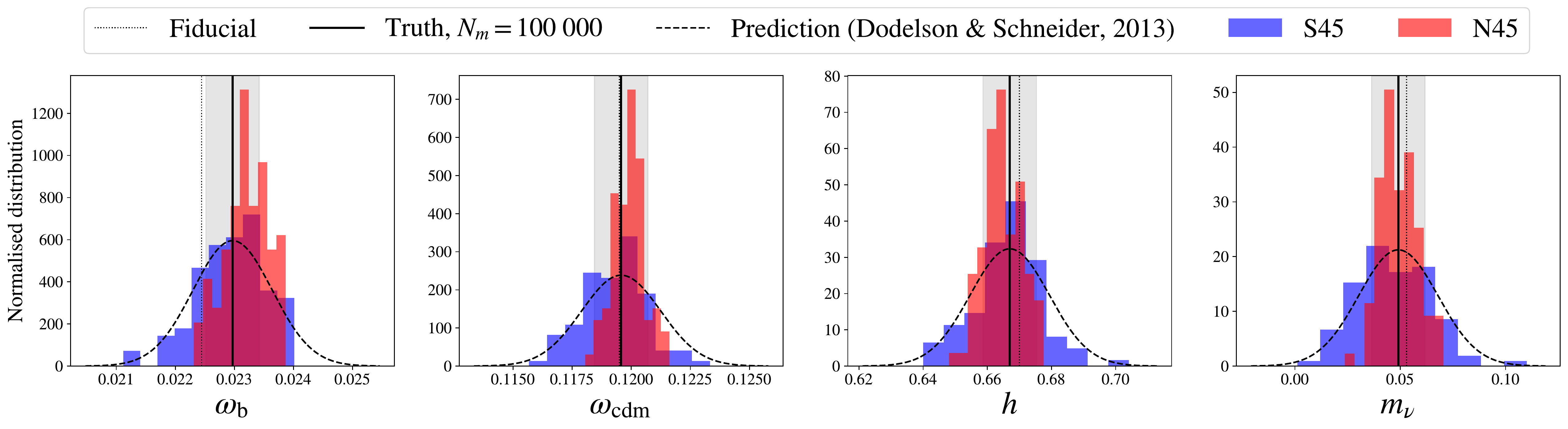}
        \caption{Distribution of the best-fit, over $\sim 100$ fits, for each cosmological parameters. Each fit have been run with a different realisation of the covariance matrix, estimated with 45 mock catalogues, using either the standard estimator (blue) or \nercome\ (orange). The black line and area show the best-fit and error estimated with the true covariance matrix. The dashed black curve gives the prediction from \citet{dodelson_13} (c.f. Eq.~\eqref{eq:bf_disp}). The dotted black line indicates the true cosmology.}
        \label{fig:histoBF}

        \includegraphics[scale=0.35]{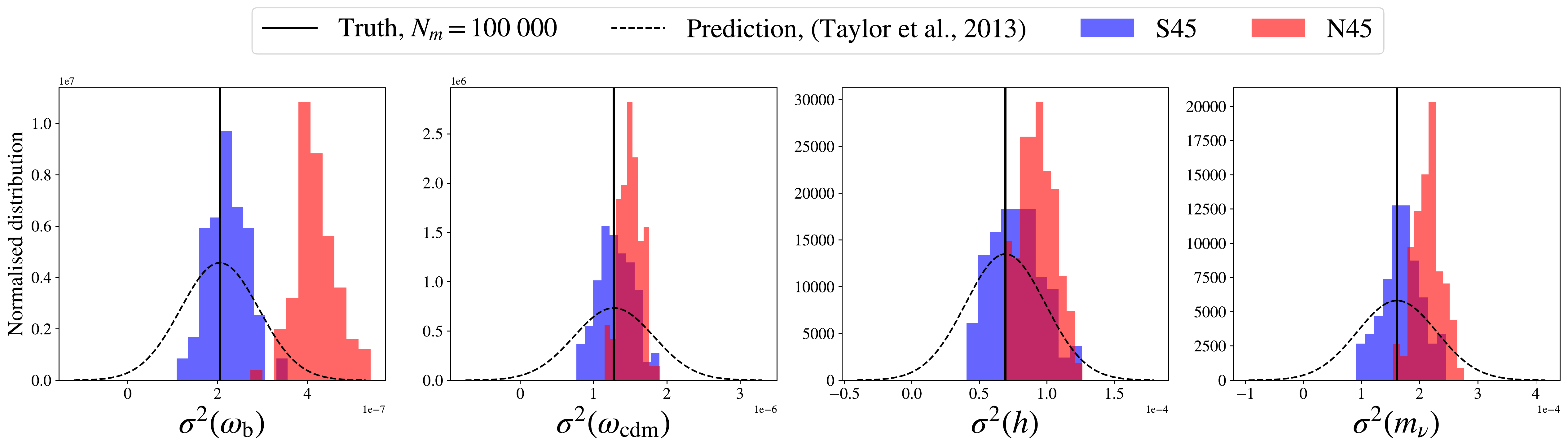}
        \caption{Same as figure~\ref{fig:histoBF} for the variance. The black line shows the variance estimated with the true covariance matrix. The dashed black curve gives the prediction from \citet{taylor_13} (c.f. Eq.~\eqref{eq:taylor_12}).}
        \label{fig:histoVar}
    \end{figure*}
    
    We have explicitly shown the effect of the Hartlap bias on parameter inference, and how the sampling noise present in the precision matrix can drastically impact the constraints on cosmological parameters. In the next sections we will show how \nercome\ and \covmos\ can mitigate these two effects.
    
    \subsection{Reduction of sampling noise with \nercome}\label{subsect:param_nercome}
 
 The aim of this section is to test whether \nercome\ can reduce the effect of sampling noise in the precision matrix, while resulting in unbiased cosmological constraints. To isolate the effects coming from sampling noise, it is better if the parameter inference is not affected by biases due to the modeling of the non-linear power spectrum, or to the fact that the covariance matrix is not perfectly describing the data. To overcome these issues, in this section, we chose to use the \covhf\ data set both for the data vector and the covariance matrix. In this way we make sure we  have the exact modeling and covariance for our data.
    
    \subsubsection{Performance at low $N_m$}
    
    A way to estimate the dispersion of the best-fit and error of cosmological parameters due to sampling noise, is to run several fits on the same data vector for different realisations of the precision matrix. From the 100~000 \covhf\ power spectra we can create 100 different subsets, each containing $N_m=45$ realisations. The precision matrix is estimated on each of these subsets, with the Hartlap-corrected estimator (Eq.~\eqref{eq:unbiased_psi}, dubbed standard estimator in the following ) and \nercome.

    \begin{table*}
        \centering
        \caption{Summary of statistics on the best-fit of each cosmological parameter, over all fits. The three rows on the top present the dispersion of the best-fit relatively to the error estimated with the true covariance and the reduction of the best-fit variance thanks to \nercome. The two bottom rows present the deviation of the averaged best-fit over all selected fits, with respect to the one estimated with the true covariance. This is expressed in fraction of the true error on the parameter.}
    %\begin{adjustbox}
    %\renewcommand{\arraystretch}{1.2}
    %\setlength{\tabcolsep}{10pt}
        %\bgroup
        \def\arraystretch{1.3}
        \begin{tabular}{l | c c c c c}
            \hline
            \multicolumn{1}{l |}{} & $\omegab$ & $\omegac$ & $h$ & $m_\nu$ \\
            \hline
            \multicolumn{1}{l |}{Best-fit dispersion for \texttt{S45}: $\sqrt{\mr{Var}\left[\hat{\theta}\right]}/\sigma_{\theta}$ [\%]} & 129.6 & 123.5 & 129.7 & 138.9 \\
            \multicolumn{1}{l |}{Best-fit dispersion for \texttt{N45}: $\sqrt{\mr{Var}\left[\hat{\theta}\right]}/\sigma_{\theta}$ [\%]} & 82.9 & 60.9 & 71.6 & 67.1 \\
            
            \multicolumn{1}{l |}{Reduction of the best-fit dispersion: \texttt{N45} / \texttt{S45} - 1 [\%]} & -36.0 & -50.7 & -44.8 & -51.7 \\
            \hline
            \multicolumn{1}{l |}{Mean best-fit shift wrt truth for \texttt{S45}: $\left(\langle \hat{\theta}\rangle-\theta\right)/\sigma_{\theta}$} & $<|0.1|$ & $-0.2$ & $<|0.1|$ & -0.1  \\
            \multicolumn{1}{l |}{Mean best-fit shift wrt truth for \texttt{N45}: $\left(\langle \hat{\theta}\rangle-\theta\right)/\sigma_{\theta}$} & 0.5 & 0.3 & -0.3 & $<|0.1|$ \\
        \end{tabular}
        \label{tab:bfdisp}
    \end{table*}
    
    \vspace{1cm}
    \begin{table*}
    \centering
    \caption{Summary of statistics on the variance of each cosmological parameter, over all fits. The three rows on the top present the dispersion of the variance relatively to the variance estimated with the true covariance and the reduction of this dispersion thanks to \nercome. The two bottom rows present the relative difference between the averaged variance over all selected fits, with respect to the true variance.}
    \label{tab:vardisp}
    %\begin{adjustbox}
    %\renewcommand{\arraystretch}{1.2}
    %\setlength{\tabcolsep}{10pt}
    %\bgroup
    \def\arraystretch{1.3}
    \begin{tabular}{l | c c c c c}
        \hline
        \multicolumn{1}{c |}{} & $\omegab$ & $\omegac$ & $h$ & $m_\nu$ \\
        \hline
        \multicolumn{1}{l |}{Variance dispersion for \texttt{S45}: $\sqrt{\mr{Var}\left[\hat{\sigma}^2_\theta\right]}/\sigma^2_{\theta}$ [\%]} & 22.4 & 18.8 & 27.7 & 20.7 \\
        \multicolumn{1}{l |}{Variance dispersion for \texttt{N45}: $\sqrt{\mr{Var}\left[\hat{\sigma}^2_\theta\right]}/\sigma^2_{\theta}$ [\%]} & 22.3 & 11.8 & 18.6 & 14.6 \\

        \multicolumn{1}{l |}{Reduction of the variance dispersion: \texttt{N45} / \texttt{S45}-1 [\%]} & -0.5 & -37.2 & -33.0 & -29.5  \\
        \hline
        \multicolumn{1}{l |}{Mean variance shift wrt truth for \texttt{S45}: $\langle \hat{\sigma}^2_\theta \rangle/\sigma^2_{\theta}-1$  [\%]} & 8.7 & 2.0 & 12.4 & 5.4 \\
        \multicolumn{1}{l |}{Mean variance shift wrt truth for \texttt{S45}: $\langle \hat{\sigma}^2_\theta \rangle/\sigma^2_{\theta}-1$  [\%]} & 104.9 & 17.7 & 34.2 & 32.6 \\
        \hline 

        \end{tabular}
    \end{table*}
    
    We perform 100 MCMC's with the data-vector being one realisation of the \covhf\ set (for each of the 5 redshifts) for a fixed $\kmax=0.2\ \invMpc$ (corresponding to $N_b = 30$), with the 2 different estimations of the covariance matrix:
    \begin{itemize}
        \item standard estimator, with $N_m=45$, dubbed \texttt{S45},
        \item \nercome\ with $N_m=45$, dubbed \texttt{N45}.
    \end{itemize}
    Note that in the \texttt{S45} case we don't apply the $m_1$ correction to the resulting constraints in order to gauge how \nercome\ compares with the "raw" constraints.

    We have then access to the distribution of best-fits and errors for the $4$ fitted cosmological parameters in the $2$ cases. As a reference, we also perform a fit with the covariance matrix estimated with $N_m = 100~000$, which can be considered as the true covariance. All this is done in the 16$\nu$ cosmology so that $m_\nu$ is well constrained and we only deal with Gaussian posterior distributions (c.f. Figure~\ref{fig:triangle_high_samnoise}).
    
    In Figure~\ref{fig:histoBF}, we examine the distribution of the best-fit for \texttt{S45} and \texttt{N45}, compared to the best-fit obtained using the true covariance. Table~\ref{tab:bfdisp} gathers the statistics of these distributions. First we can see that the fiducial values of the cosmological parameters are recovered within, or very close to $1\sigma$ in the case of the true covariance matrix. Then, the dispersion of the best-fit is of the same order of magnitude than the true error, for both \texttt{S45} and \texttt{N45}. Still, \nercome\ reduces this dispersion from around $130$ to $70$\% of the true error, which corresponds to a decrease between $\sim 35$ and $50\%$ depending on the parameter. This dispersion is stochastic, so the best-fit should not deviate from the truth in a preferred direction. Looking at the bottom of Table~\ref{tab:bfdisp} we see that indeed the average deviation of the best-fit with respect to the truth is not statistically significant. Finally, we can see that the prediction from \citet{dodelson_13} (Eq.~\eqref{eq:bf_disp}) is in quite good agreement with the distributions for \texttt{S45} shown in Figure~\ref{fig:histoBF}. We will present a more detailed comparison in the next section.
    
    After analysing the effect of sampling noise on the best-fit, let us focus on the variance of each parameter, estimated from the marginalised posterior distribution. Figure~\ref{fig:histoVar} shows the distribution of the variance, $\sigma^2_\theta$, for \texttt{S45} and \texttt{N45}, compared to the variance estimated using the true covariance. The statistics of these distributions are exposed in Table~\ref{tab:vardisp}. In the \texttt{S45} case, the dispersion of the parameter variances correspond to about 25\% of the true variance for all cosmological parameters. With the case of \texttt{N45}, we estimate that \nercome\ reduces the variance on the variance by 30\% for all parameters except $\omegab$ for which there is no reduction. In the same time we observe a systematic overestimation of the parameter variances with \nercome. Indeed, when using a covariance matrix estimated with \nercome, the cosmological parameters variances are larger than the true ones by about 17\% for $\omegac$, 30\% for $h$ and $m_\nu$ and even going up to 100\% for $\omegab$, while this bias is only between 2 and 12\% for \texttt{S45}. So, despite allowing a decrease of sampling noise effects, \nercome\ induces a significant bias on the variance of cosmological parameters. Finally, we see in Figure~\ref{fig:histoVar} that the prediction from \citet{taylor_13} (Eq.~\eqref{eq:taylor_12}) does not match the estimated distribution of the variances. This will be further discussed in the next section.
    
    \subsubsection{Evolution with $N_m$ }

\begin{figure*}
        \centering
        \includegraphics[scale=0.4]{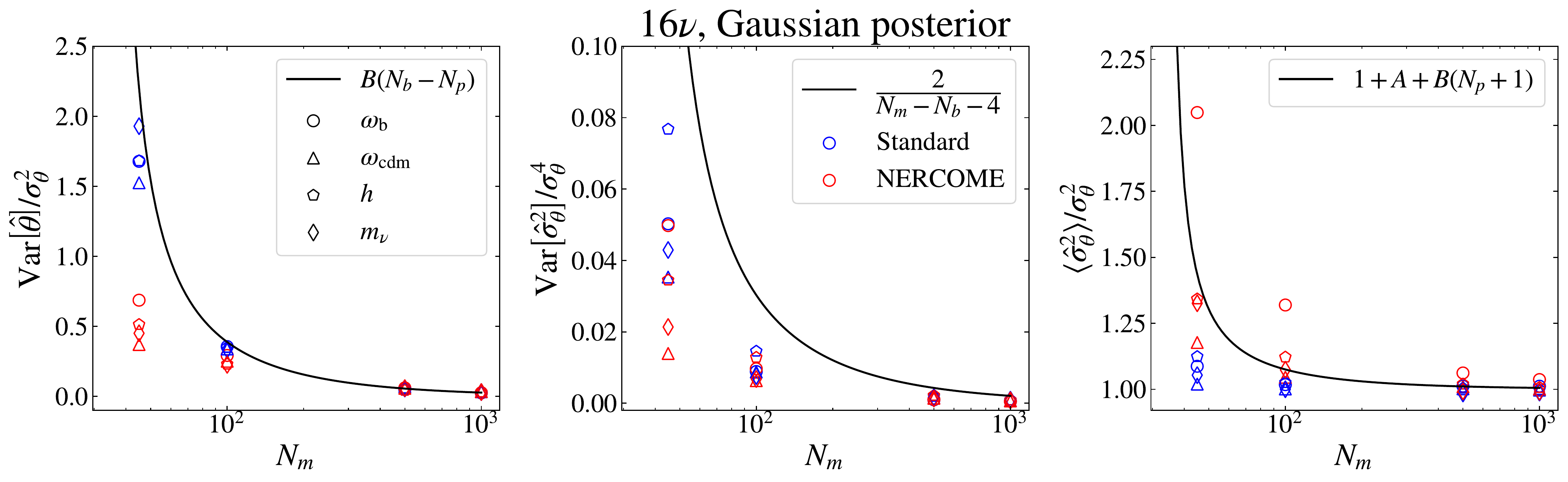}
        \caption{Comparison of the prediction of sampling noise effects with the estimation on $\sim$100 fits using a covariance matrix estimated either with the standard estimator or \nercome, for $N_m=45,\ 100,\ 500$ and 1000. The results are shown for the $16\nu$ cosmology, resulting in a Gaussian posterior distribution. \textit{Left}: ratio of the variance on the best-fit to the true variance, compared to \citet{dodelson_13} (Eq.~\eqref{eq:bf_disp}). \textit{Middle}: ratio of the variance on the variance to the true variance squared, compared to \citet{taylor_13}(Eq.~\eqref{eq:taylor_12}). \textit{Right}: ratio of the mean variance to the true variance compared to \citet{percival_13}(Eq.~\eqref{eq:var_bias}).}
        \label{fig:compare_pred_16nu}
    \end{figure*}

    In this section we have done the same exercise as above but for several values of $N_m$, still in the 16$\nu$ cosmology. Figure~\ref{fig:compare_pred_16nu} shows the evolution with $N_m$ of the three sampling noise effects on the estimated parameters, which are the dispersion on the best-fit ${\rm Var}[\hat\theta]$, the variance on the variance ${\rm Var}[\hat\sigma_\theta^2]$ and the bias on the variance $\langle\hat\sigma_\theta^2\rangle$, compared to their respective analytic predictions. With our 100~000 realisations of measurements of the power spectrum we make subsets of up to $N_m=1000$ realisation in order to obtain $100$ estimations of the precision matrix. We keep $N_b=30$ as in the previous sub-section. 
    
    %We show this when using a covariance matrix estimated either with the standard estimator or NERCOME.
    
    In the case of the standard estimator, the dispersion on the best-fit (left panel of Figure \ref{fig:compare_pred_16nu}) is following the analytical prediction by \citet{dodelson_13}. In addition, we clearly see the effect of \nercome\  which reduces this dispersion when the number of realisation $N_m$ used to estimate the precision matrix is low. However, the best-fit dispersion obtained with \nercome\  converges toward the one expected with the standard estimator for $N_m \sim 100$. In this limit the variance of the best fit is lower than half the variance on the fitted parameters. 
    
    Regarding the variance of the variance of the estimated parameters, we see that \nercome\  is only slightly reducing it and also converges towards the standard estimator for $N_m = 100$. One can see that even when $N_m$ is increasing the dispersion of the variance of the fitted parameters is systematically lower (roughly a factor of $2$) than expected by \citet{taylor_13}. This is in good agreement with what has been noticed in the previous subsection for $N_m = 45$. Anyway, the variance on the variance of estimated cosmological parameters is only contributing at a level lower than $10$\% even for $N_m=45$. This contribution drops around $1$\% for $N_m=100$ which makes it the least dominant contribution to the total error budget. 
    
    In turn, for what concerns the bias on the variance at low $N_m$ we can clearly see again that despite the reduction of the two effects discussed above, the use of \nercome\  results in an increase of the parameter variance. This increase becomes negligible for $N_m = 1000$ but is still slightly present for $\omegab$ at $N_m = 500$. Considering the standard estimator one can see that the bias is much smaller than what predicts \citet{percival_13}. Indeed, the estimated uncertainties on cosmological parameters is only of the order of $10$\% larger than the true one when using the standard estimator with only $45$ realisations. 

    While we find the predictions from \citet{taylor_13} and \citet{percival_13} to disagree with our estimations, it is important to note that they correspond to small effects and that at least they appear to be conservative. 
    
    In the end, this shows that for what concerns the error on cosmological parameters it seems equivalent to use the standard covariance estimator and to inflate error-bars by $\sqrt{m_1}$ or to use \nercome. Indeed, the bias on the variance of parameters we get when using \nercome\ is about the same order as the best-fit dispersion we get when using the standard estimator (which is what $\sqrt{m_1}$ corrects for).
    
    %This is indeed confirming that using the standard estimator for $N_m=100$ one would estimate the right error on cosmological parameters but with a variance on the best fit of about $40$\% of the true variance this would result in the necessity of artificially increasing the effective error on cosmological parameters so to avoid any possible bias in the cosmological inference. To do so one should add in quadrature the expected dispersion on the best fit and the estimated error on cosmological parameters. 
    
    In conclusion, although \nercome\ is quite efficient in reducing the dispersion on the best-fit in a low $N_m$ regime, the resulting cosmological errors are biased. Thus, \nercome\ might be used to gauge by how much the constraints one gets from an analysis are affected by the dispersion of the best-fit and to get a more precise idea of the best-fit position, but we cannot trust the resulting error bars.

    \begin{figure*}[h]
        \centering
        \includegraphics[scale=0.4]{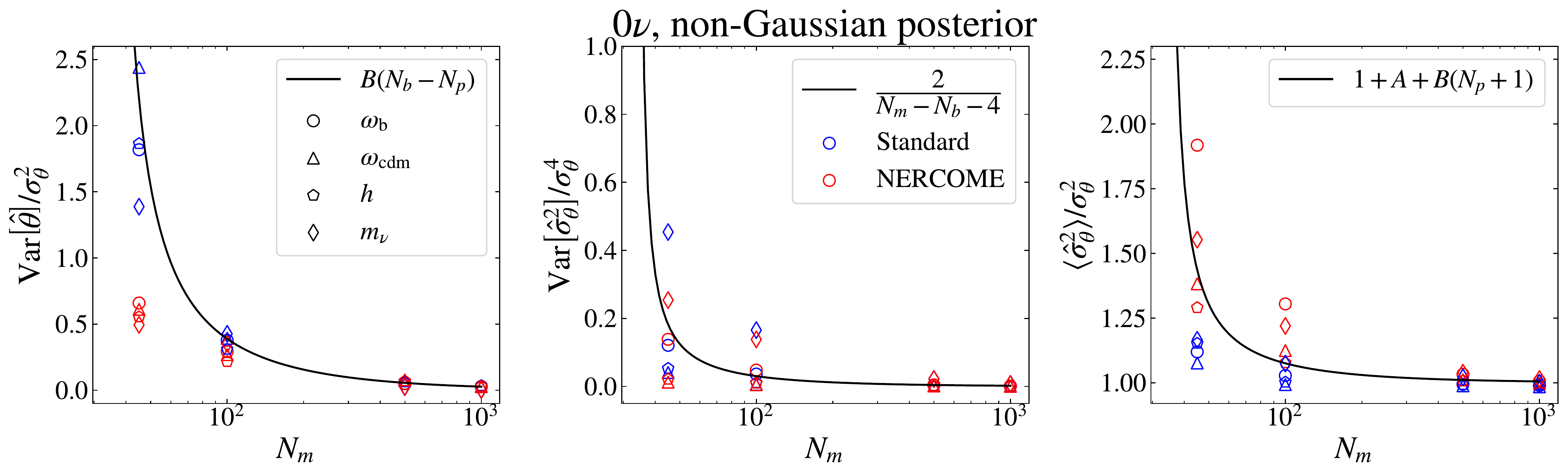}
        \caption{Same as figure~\ref{fig:compare_pred_16nu}, but for the 0$\nu$ cosmology, resulting in non-Gaussian posterior distribution.}
        \label{fig:compare_pred_0nu}
    \end{figure*}

    \subsubsection{Non-Gaussian posterior}
    
    One final test we want to conduct is to study the impact of a non-Gaussian posterior distribution on the performance of \nercome\ and on sampling noise effects in general. To do so we perform the same analysis as in the previous section but in the 0$\nu$ cosmology. In this case, as the fiducial value of the neutrino mass is $m_\nu = 0$, the estimated posterior distribution is cut by the physical prior $m_\nu >0$ (c.f. Appendix~\ref{app:ng_post_0nu} for an example). In consequence the main assumption of the analytic predictions is broken. In that case, as in the Gaussian posterior case, we take the diagonal of the parameter covariance matrix, estimated from the MCMC, to be the variance of cosmological parameters to which we compare the analytical predictions. We could also have compared the predictions with the asymmetric 68\% confidence interval as it is more representative of the uncertainty on a parameter, but the quantity that is predicted here is the diagonal of the parameter covariance. So we stick to the former choice for consistency of the comparison.
    
    Figure~\ref{fig:compare_pred_0nu} presents the same result as Figure~\ref{fig:compare_pred_16nu} but for the $0\nu$ cosmology. For the dispersion on the best-fit and the bias on the variance (left and right panels) we observe the same behavior as in the case of Gaussian posteriors. However the variance on the variance (middle panel) is larger at low $N_m$, especially for $m_\nu$ which has a non-Gaussian posterior. Moreover, in that case \nercome\  is quite efficient in mitigating this effect as it reduces the variance on the variance of $m_\nu$ by a factor $2$. Despite this increase of the dispersion of the variance one should apply the same procedure as in the Gaussian case in order to be more conservative. 

    As a last comment, from both Figure~\ref{fig:compare_pred_16nu} and Figure~\ref{fig:compare_pred_0nu}, we remark a certain scattering between the cosmological parameters. It might come from the different correlations among the parameters making them react in different ways to covariance effects. Still, we find the same order of magnitude from one parameter to another.

        %\syl{Two points related to this section could be studied in more details:
        %\begin{itemize}
        %    \item Effect of non-Gaussian posteriors. Can be studied with fit of the $0\nu$ cosmology. Fits are currently running.
        %    \item Why predictions of variance and best-fit dispersion do not agree ? Non-Gaussianity in the likelihood, effect of priors ??
        %\end{itemize}}
        
        %\sylcol{In the previous section, only show the case with $N_m = 45$, to focus on the reduction of sampling noise with NERCOME. In this section, focus on the comparison with analytic prediction. Show the case with $N_m=1000$ (maybe only in a table).}
    
    \subsection{Reliability of \covmos\ for parameter inference }\label{subsect:param_covmos}
    
    Thanks to the efficiency of \covmos\ to produce a high number of realisations, the effect of sampling noise in the precision matrix can be reduced to a negligible amount. However, it is not guaranteed that the method results in an unbiased covariance. Indeed, it was shown in \citet{baratta_22} that at $z=0$, \covmos\ over-predicts non-diagonal elements of the real space covariance for $k>0.17h/$Mpc. It is therefore important to asses whether these deviations, in the \covmos\ covariance with respect to the accurate (but noisy) \demcov\ covariance, bias the estimation of cosmological parameters.
    
    If we compare the results obtained from a \covmos\ covariance with the one obtained from a \demcov\ covariance estimated with the standard estimator, it will be hard to draw conclusions given the large influence of sampling noise in the latter case due to the low number of realisations (c.f. section~\ref{subsect:param_inf_noise}), especially on the best-fit. Thus we also include the \demcov\ covariance estimated with \nercome, as it was shown to result in reduced best-fit dispersion in the previous section. This will also allow to compare in a common setting the result of the two approaches, \nercome\ and \covmos, at the level of the parameters best-fit position.  However, it was observed in section~\ref{subsect:param_nercome}  that \nercome\ leads to bias in the estimated variance of parameters. Hence we will not be able to make a rigorous comparison at the level of parameters error-bar. 
    
    \begin{figure*}[h!]
    \centering
    \includegraphics[scale=0.65]{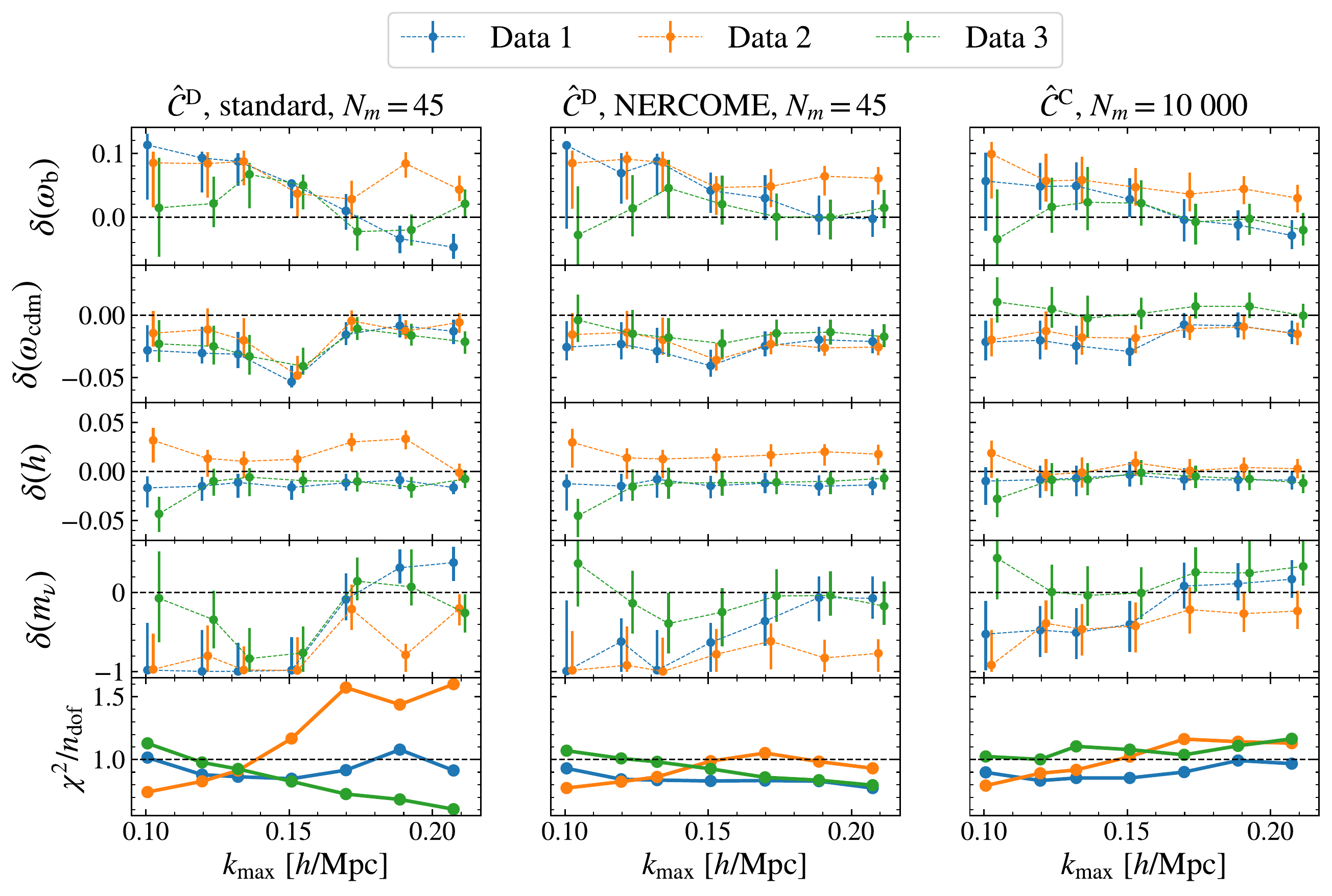}
    \caption{Parameter constraint with respect to $\kmax$, for 3 different sets of \demcov\ realisations: \texttt{DEMNUni-Cov01-05} in blue, \texttt{DEMNUni-Cov06-10} in orange, \texttt{DEMNUni-Cov46-50} in green. The three columns correspond to the covariance matrix which was used in the fits: standard \demcov\ covariance with $N_m=45$ and corrected for the Hartlap bias (left), \nercome\ \demcov\ covariance with $N_m=45$ (middle) and standard \covmos\ covariance with $N_m=10~000$ (right). The top rows show the relative difference with respect to the fiducial cosmology for each parameter and the bottom row the $\chi^2/n_\mr{dof}$. }
    \label{fig:maxi_param}
    \end{figure*}
    
    For this exercise we will use the \demcov\ power spectrum as our data-vector (still at the 5 redshifts). We make this choice because we want to test the capability of \covmos\ to provide an unbiased parameter estimation in a realistic case. Here we mean realistic in the way that \covmos\ should be used, {\it i.e.} cloning an existing data-set to enlarge its number of realisations. In other words, we test, at the level of the estimated parameters, how well \covmos\ can reproduce the \demcov\ simulations\footnote{The validation has been made at the level of the power spectrum, the 2-pt correlation function and their covariance matrices in \citet{baratta_22}.}.
    
    To account for the intrinsic noise in the data vector, the fits are done for three different data vectors, \texttt{Data1}, \texttt{Data2} and \texttt{Data3}, each containing one independent realisation of the power spectrum per redshift, taken from the 50 \demcov. Note that we performed the same analysis for the 10 possible independent combinations of data, but we only show three of them for clarity because it does not change the conclusions.
    
    For the covariance matrix, as explained above we consider three cases:
    \begin{itemize}
        \item $\hat{\mbf{C}}^\mr{D}$, standard. A covariance matrix estimated with the standard estimator from the $N_m = 4$5 remaining \demcov\ realisations. We correct for the Hartlap bias in the precision matrix but we don't apply the $m_1$ factor to the resulting constraints.
        \item $\hat{\mbf{C}}^\mr{D}$, \nercome. Same as above but with the \nercome\ estimator.
        \item $\hat{\mbf{C}}^\mr{C}$. A covariance matrix estimated from the $N_m = 10~000$ \covdem\ realisations (c.f. section~\ref{subsect:covmos}).
    \end{itemize}
    
    Finally, to isolate the effect of the covariance, we do not consider $\kmax > 0.2\ \invMpc$ to limit the impact of biases coming from the modeling of the non-linear power spectrum (c.f. \ref{subsect:param_inf_noise}).
    
    %Because we want to test the capability of \covmos\ to provide an unbiased parameter estimation in a realistic case, the fits are carried out on the \demcov\ power spectrum (not on the \covhf\ ones, as in the previous section).
    
    \begin{figure*}[h!]
    \centering
        \includegraphics[width=60mm]{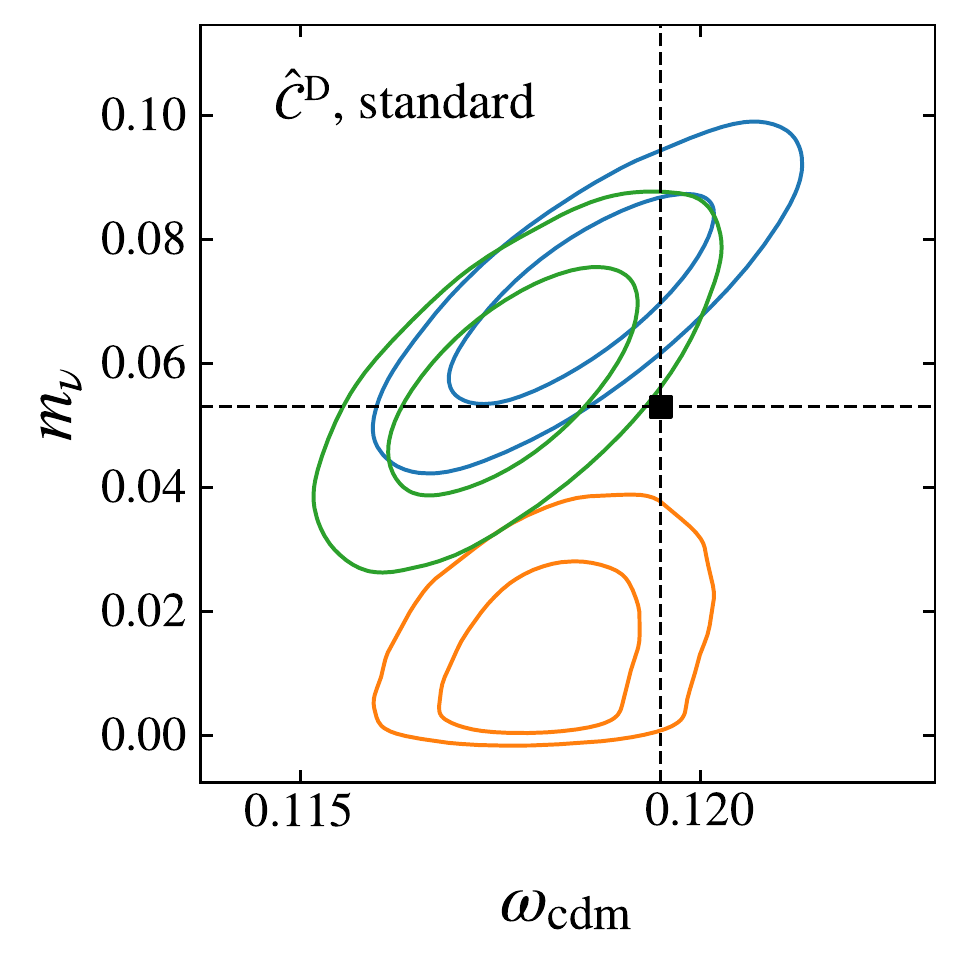}
        \includegraphics[width=60mm]{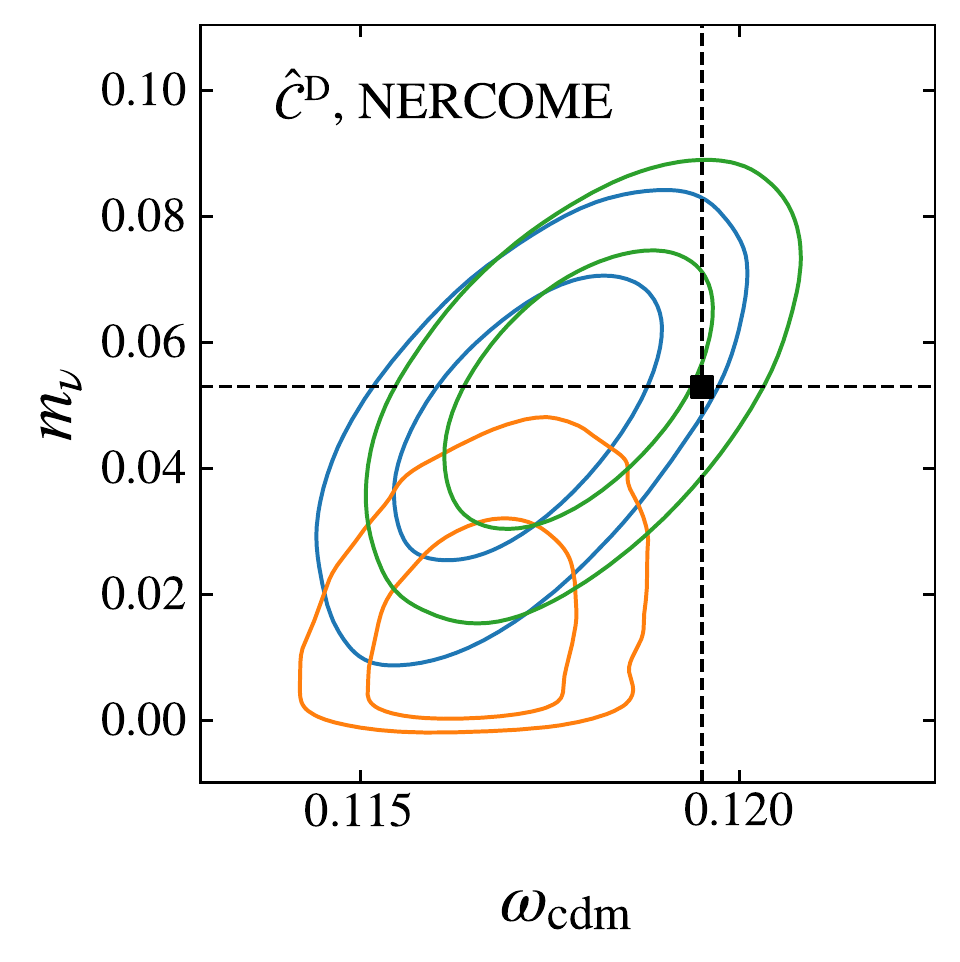}
        \includegraphics[width=60mm]{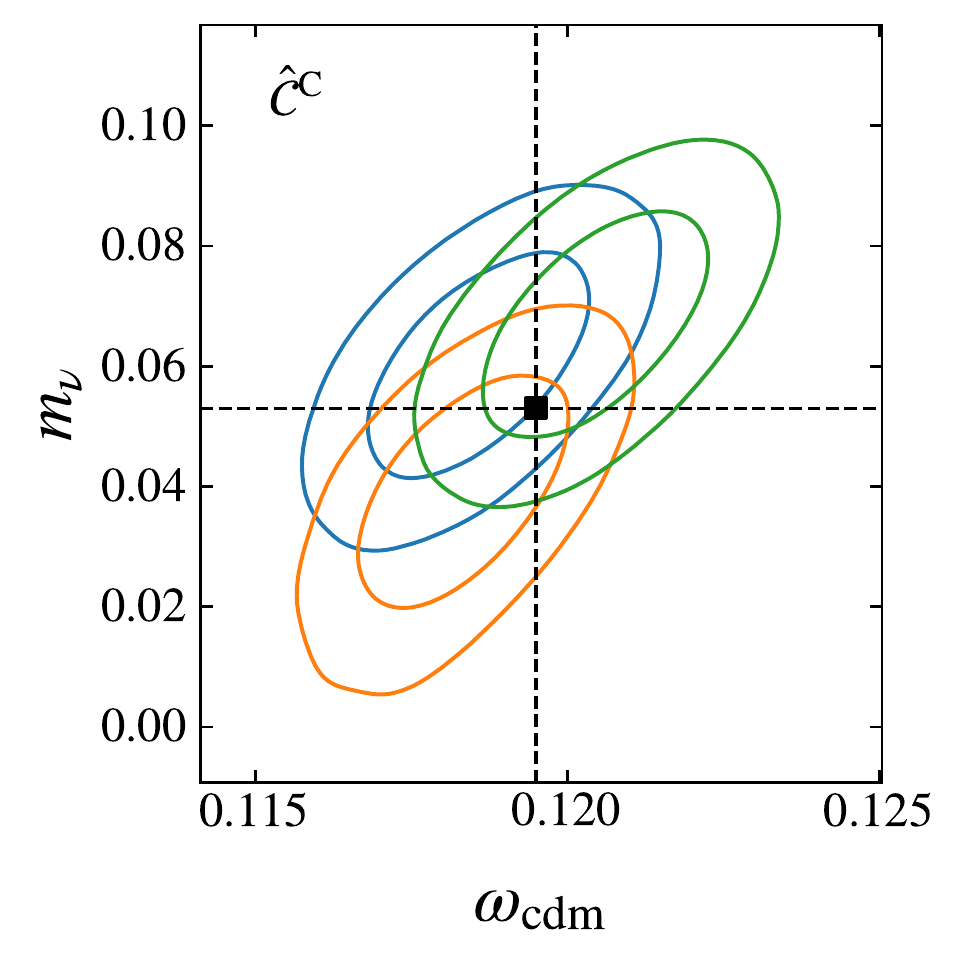}
        \caption{2D marginalised posteriors in the ($\omegac$, $m_\nu$) plan, for $\kmax = 0.19\ \invMpc$. The colors correspond to the %3 
        three different data sets as in Figure~\ref{fig:maxi_param}. The 
        three panels correspond to the different covariances considered. The black dashed lines and the square show the fiducial cosmology.}
    \label{fig:3_2d}
    \end{figure*}

    \begin{figure*}[h]
        \centering
        \includegraphics[scale=0.6]{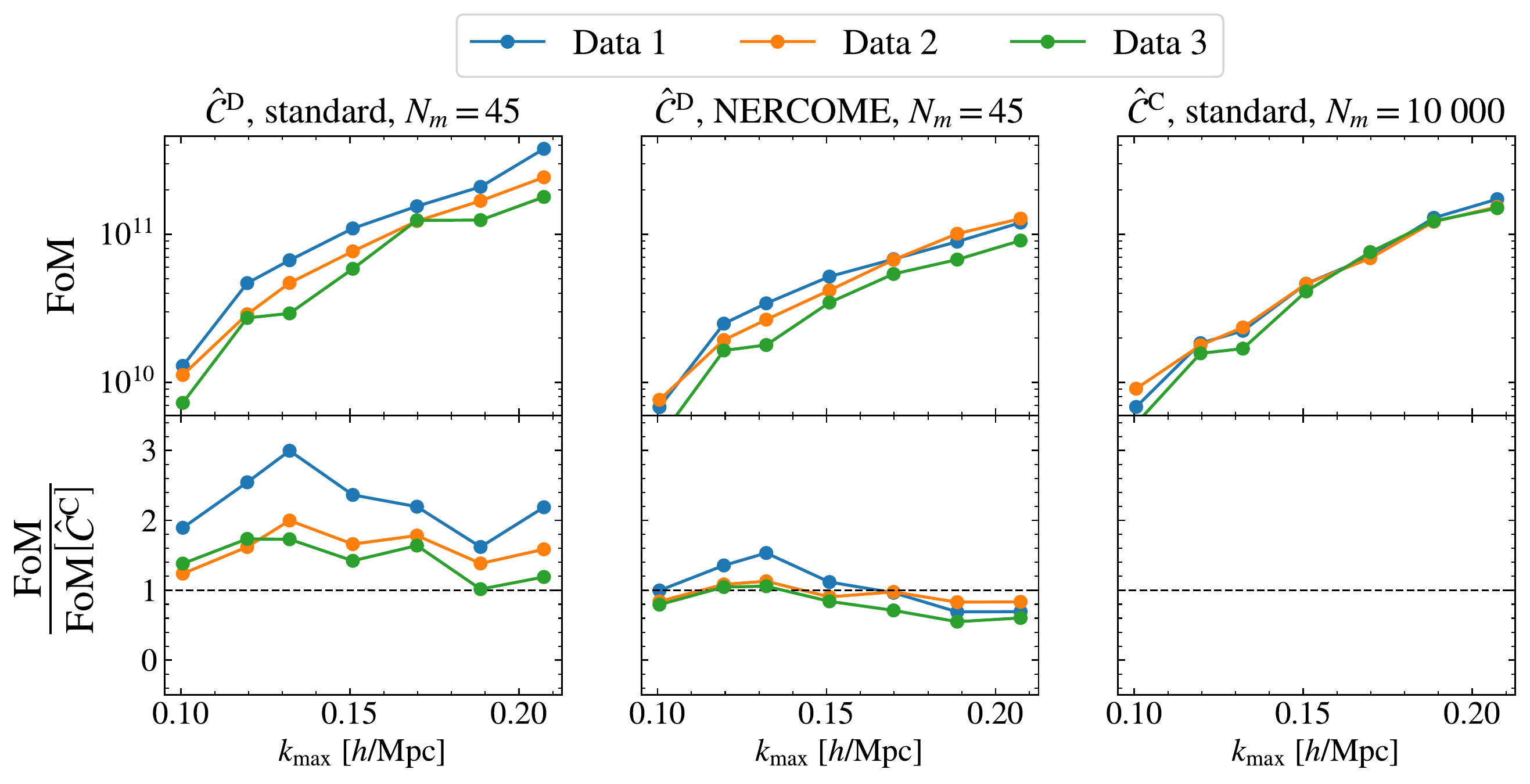}
        \caption{The top panel shows the FoM for the 3 different \demcov. The bottom panel shows the ratio of the FoM obtained with the standard ({\it left}) and \nercome\ ({\it middle}) estimation of the \demcov\ covariance, to the FoM obtained with \covmos\ ({\it right}).}
        \label{fig:maxi_fom}
    \end{figure*}
    
    Figure~\ref{fig:maxi_param} displays the 1D marginalised posteriors of cosmological parameters as well as the $\chi^2/n_\mr{dof}$ for $\kmax$ between $0.1\ \invMpc$ and $0.2\ \invMpc$, for the three data vectors and the three types of covariance matrices we consider. To understand this figure we summarise the relevant information in the following bullet-points:
    \begin{itemize}
        \item Some points feature asymmetric error-bars, indicating that the posteriors are not Gaussian, especially at low $k$. This is mostly due to the fact that, in some cases the estimation of $m_\nu$ is compatible with 0, forcing the posterior to be cut at the low boundary of the prior on this parameter.
        \item At $\kmax=0.1\ \invMpc$, for all parameters, and covariances, the scatter on the best-fit, for the different data-sets is larger than for higher $\kmax$. This comes from the intrinsic variance of the power spectrum which is larger on large scales and is also reflected in the parameter error-bars which, are larger for this scale cut. 
        %We can see that this scatter is slightly reduced for the \covmos\ covariance, because it has a negligible noise contribution. 
        \item On the left panel, corresponding to the standard \demcov\ covariance, we can see a large scatter of the best-fit across the $\kmax$ range. The same applies for the $\chi^2$. This indicates that with such covariance, we can potentially have an estimation of cosmological parameters which greatly deviates from the truth, by more than 3$\sigma$ in some cases. It was expected and already observed in section~\ref{subsect:param_inf_noise}. 
        \item In contrast, by looking at the middle panel, the \demcov\ covariance estimated with \nercome, exhibits more stable constraints with respect to $\kmax$, thanks to the diminution of the best-fit scattering observed in the previous section. However, in some cases we see a systematic deviation from the fiducial cosmology, larger than $1\sigma$. This is present especially in the case of \texttt{Data2}, which does not agree with the two others, especially for $h$. This disagreement mainly comes from residual sampling noise in this covariance matrix.
        \item Finally, on the right panel, when the parameters are estimated, using the \covmos\ covariance matrix, the best-fits are generally more centered on the fiducial values. This can be seen especially in the case of \texttt{Data2}. Considering all the data-sets, the \covmos\ covariance gives a close to $1\sigma$ agreement for all parameters in the $\kmax$ range $[0.15,0.2]\ \invMpc$. In addition, the $\chi^2/n_\mr{dof}$ is stable and close to 1 on all cases.
    \end{itemize}

    For completeness we also show in Figure~\ref{fig:3_2d} the 2D marginalised posteriors, only in the ($\omegac$, $m_\nu$) for simplicity, for all data-sets and covariances. We clearly see the advantage of using \covmos\ here, as in the two other cases the estimated parameter posteriors present a strong bias with respect to the fiducial cosmology. In addition, we see again that with \nercome\ although the ellipses are less scattered than with the standard estimator $\hat{\mbf{C}}^\mr{D}$, their width are overestimated compared to the two other cases.  
    
    After studying the bias in the best-fit, let us focus on the errors. The metric we consider for this is the Figure of Merit (FoM), defined as 
    \be
        \mathrm{FoM} \equiv \dfrac{1}{\sqrt{\det(\hat{\bs{\Phi}})}}, 
    \ee
    where $\hat{\bs{\Phi}}$ is the estimated parameter covariance matrix. It is inversely proportional to the hyper-volume delimited by the 2$\sigma$ contours in the full parameter space. The higher is the FoM, the tighter are the constraints. In Figure~\ref{fig:maxi_fom}, we present the FoM with respect to $\kmax$, for all the considered cases:
    \begin{itemize}
        \item As expected, the FoM increases with $\kmax$, because the number of available modes in the power spectrum increases.
        \item Similarly to the best-fit, the FoM is scattered along the $\kmax$ range in the case of the standard \demcov\ covariance. This dispersion is lessen for the two other covariances. 
        \item In the case of the standard and \nercome\ estimation of $\hat{\mbf{C}}^\mr{D}$ the FoM also presents a dispersion among the three data vectors, which is less important for \covmos. This is a combined effect of both changing the data and the covariance for the first two cases\footnote{For a selection of 5 \demcov\ realisations going to the data vector, the covariance is estimated on the remaining $45$ simulations. So, when the data vector is changed, the covariance also changes.}, while the \covmos\ covariance is fixed.
        \item Despite this dispersion, the standard \demcov\ covariance generally results in a higher FoM than in the case of \covmos, while \nercome\ gives a slightly lower FoM for $\kmax>0.15\ \invMpc$, which confirms what was observed above.
    \end{itemize}

    As noted before, it is hard to draw rigorous conclusions on whether the constraints we obtain with \covmos\ present accurate errors because we are comparing it with constraints affected by sampling noise effects. However, we saw in section~\ref{subsect:nercome} that (i) sampling noise effects affecting the error-bars are subdominant and (ii) \nercome\ tends to result in overestimated error-bars. Thus, given the above observation that the FoM obtained with \covmos\ is slightly smaller (by a factor $\sim 0.5$ on average) than the one obtained with the standard $\hat{\mbf{C}}^\mr{D}$ and close to the one obtained with the \nercome\ $\hat{\mbf{C}}^\mr{D}$, it seems that \covmos\ leads to slightly overestimated error-bars. 
    
    Still, inferring such conclusions only on the basis of a FoM comparison can be a bit misleading as this metric also includes correlation between parameters, and the above argument is based on the results of section~\ref{subsect:nercome} where we were only considering the error-bars. In addition, by visually comparing the size of the ellipses of the right and left panels in figure~\ref{fig:3_2d}, the difference between the standard $\hat{\mbf{C}}^\mr{D}$ and the \covmos\ case is not that obvious. A more thorough study is required to confirm the above observation and is left for future work. 
    
    Finally, we can conclude that, thanks to \covmos\ it is possible to completely erase the effects of sampling noise, at the cost of a negligible computing time compared to $N$-body simulations. We have shown that when using a \covmos\ covariance matrix for parameter inference the constraints present no bias on the best-fit but possibly a slight overestimation of errors. It also appears to be the method offering the most stable results. Thus, in the next section we will use the \covmos\ covariances. 

\section{Impact of the non-Gaussian part of the covariance on cosmological constraints}
\label{sect:ng_cov}

After studying the impact of the precision of the covariance matrix on cosmological parameter posteriors, we want to see by how much they will be affected by a lack of accuracy in the covariance. In particular we quantify the bias on the best-fit value and error of parameters when wrongly neglecting the non-Gaussian contribution to the power spectrum covariance ({\it i.e.} $\mbf{C}^\mr{NG}$ in Eq.~\eqref{eq:pkcov}).

%Now that we have tested the validity of the covariance matrix, estimated with \covmos, we can quantify the impact of the non-Gaussian term ({\it i.e.} $\mbf{C}^\mr{NG}$ in Eq.~\eqref{eq:pkcov}) on cosmological constraints. %We saw in section~\ref{subsect:demcov} that this contribution has two effects on the covariance: it increases the variance on non-linear scales, with respect to the Gaussian approximation and it produces positive correlations between Fourier modes. This could potentially degrade cosmological constraints. The question is then: by how much ? 

\subsection{Methodology}

To study the impact of the non-Gaussian contribution, we perform MCMC fits of the power spectrum considering three different covariances:
\begin{itemize}
    \item the theoretical Gaussian covariance ({ \it i.e.} $\mbf{C}^\mr{G}$ in Eq.~\eqref{eq:pkcov}) computed with the fiducial \halofit\ power spectrum;
    \item a covariance matrix estimated from the \covmos\ realisations, but setting to zero all the off-diagonal elements, thus accounting only for the diagonal part of $\cng$;
    \item the full \covmos\ covariance matrix with all diagonal and non-diagonal elements.
\end{itemize}
In this way it is possible to asses which part of the covariance has the greatest influence on cosmological constraints.

We perform the fits in the $16\nu$ cosmology and consider the same methodology for parameter inference as described in section~\ref{sect:sim_cov_meth}, with a few modifications.

First, we choose to perform the parameter inference in the $\invMpc$ space, thus accounting for the isotropic AP effect (c.f. section~\ref{subsect:obs_space} and appendix~\ref{app:mpch}) to assess the impact of $\cng$ with more realistic parameter degeneracies.

Second, we choose the data-vector to be one realisation (per redshift) of the \covhf\ data-set. As we can perfectly predict the power spectrum of these data it allows us to be free from any bias stemming from the modeling of the power spectrum, even for $\kmax > 0.2\ \invMpc$. In this way we are capable of accurately quantify the impact of $\cng$ on the position and the width of cosmological parameters posterior distribution. From this choice it follows that the reference covariance matrix is the one estimated with $N_m = 100~000$ \covhf\ realisations.

\begin{figure}
    \centering
        \includegraphics[width=60mm]{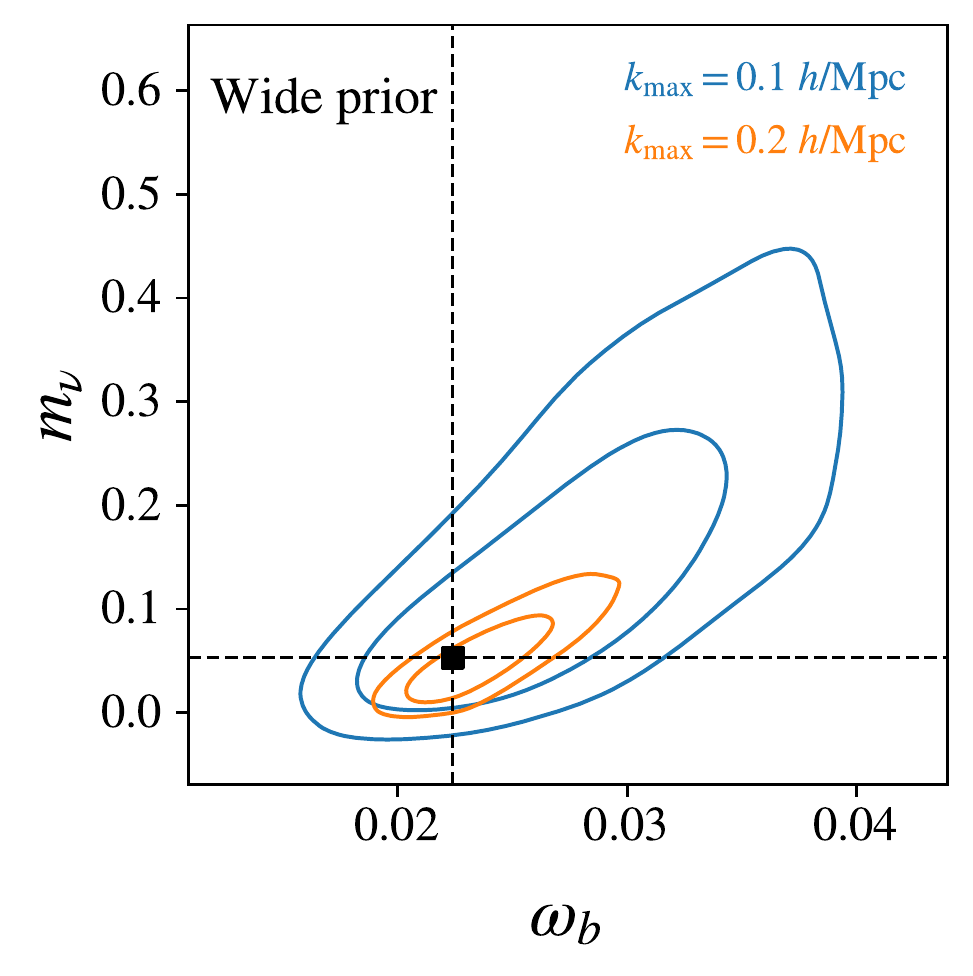}
        \includegraphics[width=60mm]{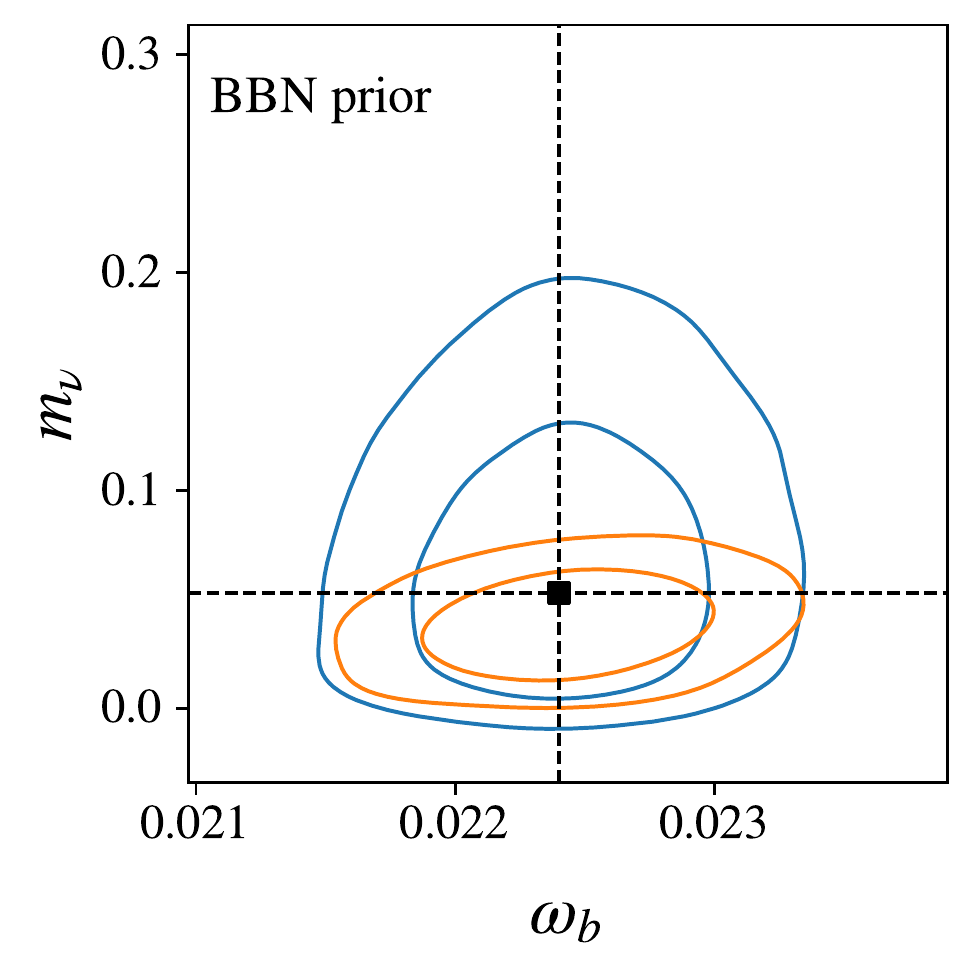}
        \caption{2D marginalised posteriors in the ($\omegab$, $m_\nu$) plan, for $\kmax = 0.1$ and $0.2\ \invMpc$. The 2 panels correspond to the type of prior considered on $\omegab$. The black dashed lines and the square show the fiducial cosmology.}
    \label{fig:prior_2d}
\end{figure}

Finally, for what regards the priors on the parameters we still consider the wide, uninformative prior described in table~\ref{tab:priors} but also, as it is usually done for $P(k)$ full-shape analysis \citep{ivanov_19}, a tighter Gaussian prior on $\omegab$ centered on the \texttt{DEMNUni} fiducial value with a width coming from Big Bang Nucleosynthesis (BBN) results \citep{cook_17} is set as
\be
\omegab = 0.0224 \pm 0.00038.
\ee
As shown in Figure~\ref{fig:prior_2d} this prior allows to break a strong degeneracy between $\omegab$ and $m_\nu$ (among other parameters). In addition this figure confirms that the fiducial cosmology is well recovered within 1$\sigma$. We don't show it to avoid an overload of information, but this is the case for all parameters and for all the scale cut considered $\kmax \in [0.1, 0.3]\ \invMpc$. The constraints discussed in this example are obtained using the full \covmos\ covariance.

\subsection{Impact of the non-Gaussian part of the covariance on parameters error}

We first quantify the impact of $\cng$ on cosmological parameters errors and we will focus on its effect on the best-fit values in the next section.

Let us define the ratio 
\begin{equation}
R(\theta) \equiv \sigma_\theta^\mathrm{NG}/\sigma_\theta^\mathrm{G}
\end{equation}
where
$\sigma_\theta^\mathrm{G}$ and $\sigma_\theta^\mathrm{NG}$ are respectively the errors on the cosmological parameter $\theta$ obtained using as covariance $\cg$ and $\cg + \cng$ ({\it i.e.} the \covmos\ covariance). 

\begin{figure*}[h]
    \centering
    \includegraphics[scale=0.65]{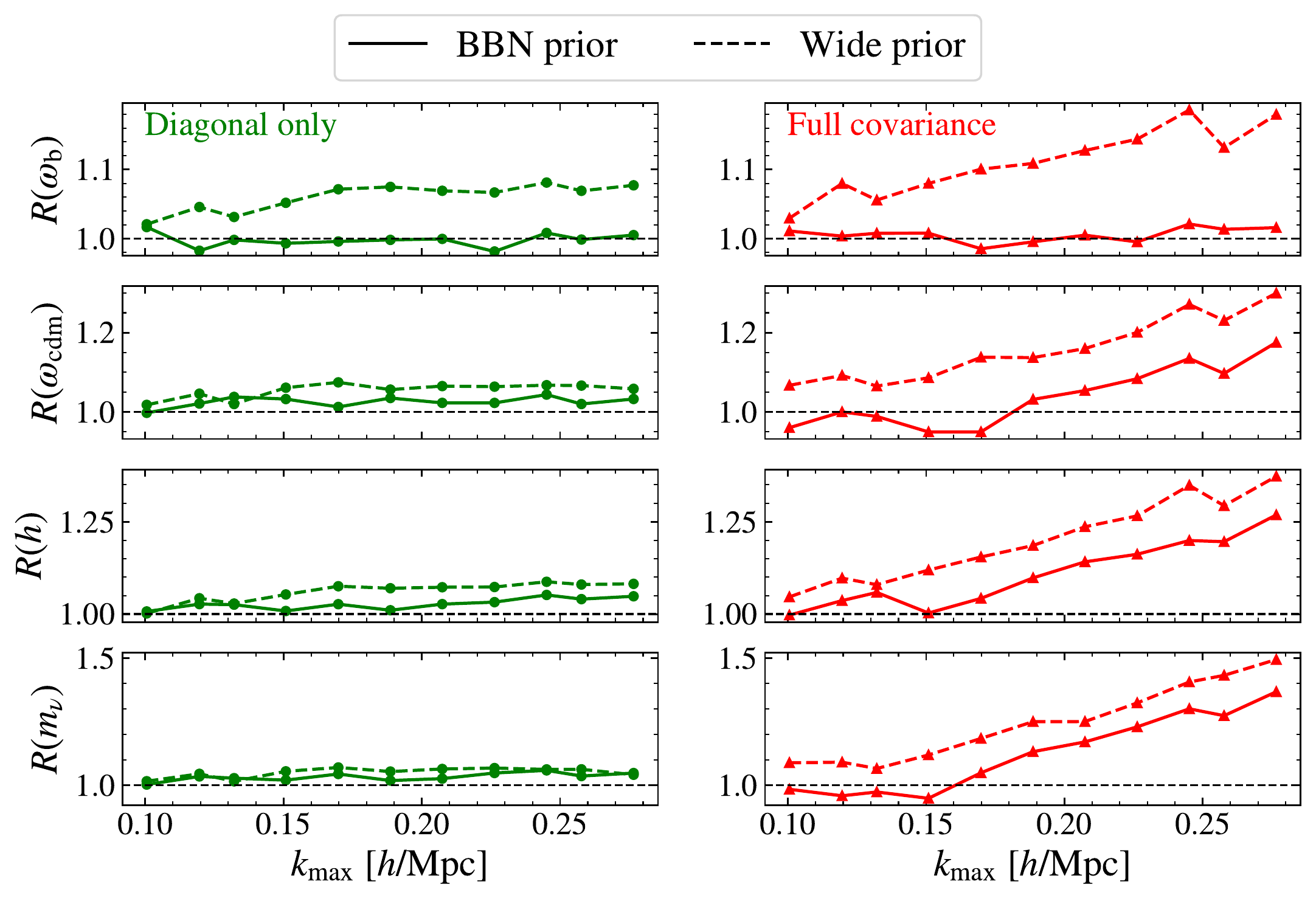}
    \caption{Relative difference in percent of marginalised cosmological parameters error between the $\cg$ only and $\cng$ case with respect to $\kmax$. In the left panel we only consider the contribution to the diagonal and in the left panel off-diagonal terms are also accounted for. The dashed and plain lines correspond respectively to the wide and BBN prior cases on $\omegab$.}
    \label{fig:err_incr}
\end{figure*}

In Figure~\ref{fig:err_incr} we show $R(\theta)$ for the whole $\kmax$ range, with or without the BBN prior and accounting or not for the off-diagonal elements of $\cng$. Without the latter and when $\omegab$ is fixed to the BBN prior the ratio is close to one for all parameters and $\kmax$. In particular, for $\omegab$, the prior is so restrictive that the error is not affected at all by the choice of the covariance. By releasing this prior the parameter uncertainty increases by 5 to 10\% depending on the parameter. Hence, the diagonal part of $\cng$ does not seem to have a large impact on cosmological constraints even for $\kmax~0.3\ \invMpc$. We recall that at these scales the total power spectrum variance is larger than the Gaussian only one by about $2\sigma$, as seen in Figure~\ref{fig:demcov_cii} with the \demcov\ covariance. In addition this is a conservative result as the \covmos\ variance we are considering here is slightly overestimating the \demcov\ one for $k>0.17~h/$Mpc as reported in \citet{baratta_22}.

When adding the off-diagonal terms $R(\theta)$ increases significantly for all parameters, except for $\omegab$ when it is restricted to the BBN prior, where it is still unaffected by the choice of the covariance. As expected $R(\theta)$ increases with $\kmax$ because the correlations between the Fourier modes come from non-linear clustering which is more important when probing small scales. In addition, this increase in $R(\theta)$ starts at $\kmax = 0.15~\invMpc$, which reflects the scale below which clustering is still linear and thus the non-Gaussian contribution to the covariance negligible.

The most affected parameter is $m_\nu$ with the error ratio in the BBN prior case going almost linearly from 1 to 1.4 between $\kmax=0.15$ to 0.3 $\invMpc$, while for $h$ and $\omegac$, the ratio reaches a maximum of 1.25 and 1.2 respectively. Without the BBN prior, $R(\omega\mr{b
})$ is at most $\sim1.2$ and all the other parameters take a $+0.1$ on $R(\theta)$ at all $\kmax$. This reflects how a prior on $\omegab$ can affect the other parameters through their correlations. In the rest of the analysis we always apply the BBN prior on $\omegab$.

We note that while $m_\nu$ is the most affected parameter it is also the least well constrained. As an example for $\kmax=0.28\ \invMpc$ in the BBN prior case and with the Gaussian covariance, the relative error ($\sigma_\theta/\theta$) on $m_\nu$ is $19\%$ while it is 1.6\%, 1.0\% and 0.4\% for $\omegab$, $\omegac$ and $h$ respectively. This is quite counter-intuitive as one would expect that the parameters most affected by a change in the covariance would be those on which the power spectrum depends the most, which are $\omegac$ and $h$ in the present case, given the aforementioned relative errors.

\subsection{Impact of the non-Gaussian part of the covariance on parameters best-fit position}

\begin{figure}[h!]
	\centering
	\includegraphics[scale=0.42]{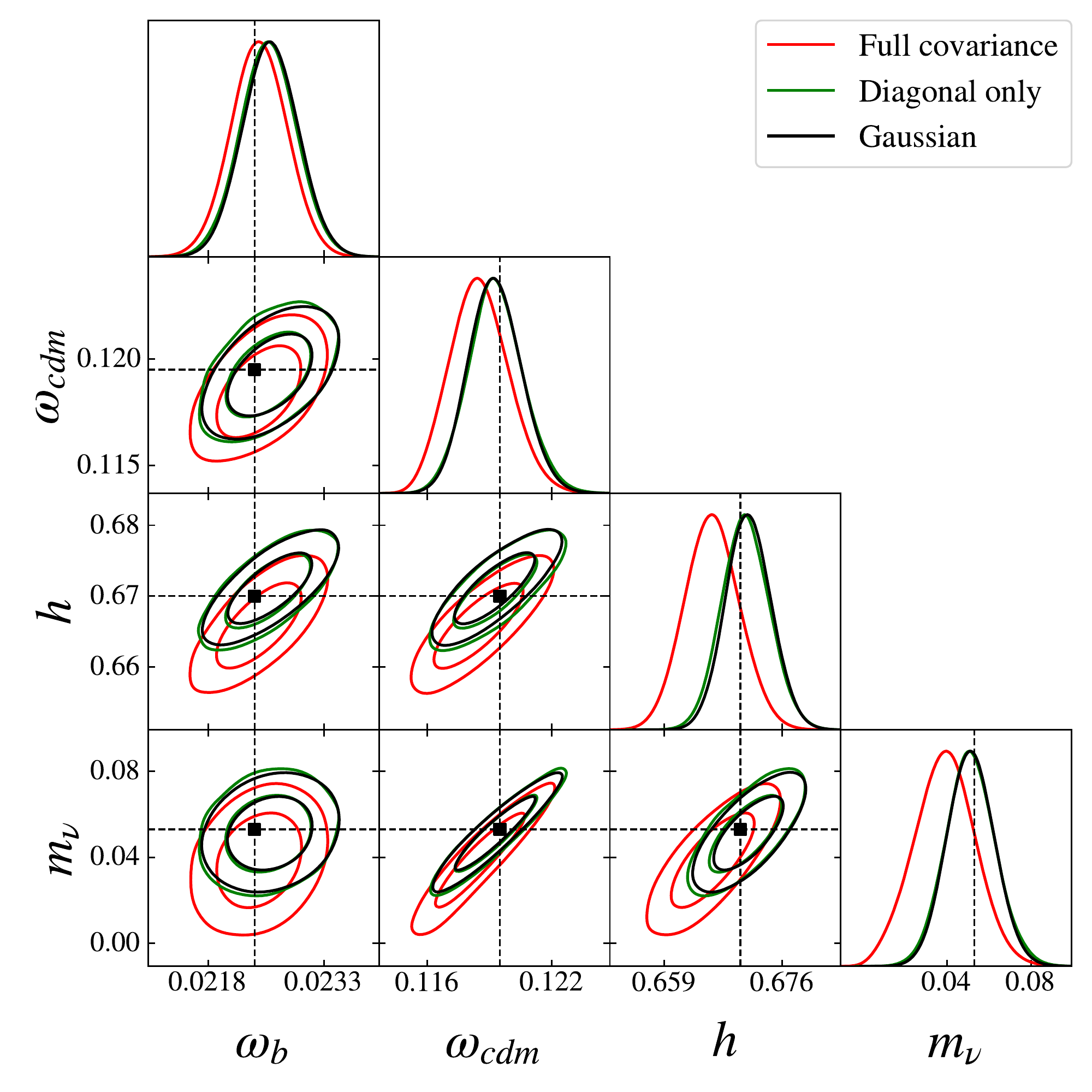}
    \caption{2D and 1D marginalised posteriors, obtained with the Gaussian covariance (black), the diagonal (green) and full (red) non-Gaussian covariance from \covmos, for $\kmax=0.25\ \invMpc$. The black square and dashed lines show the fiducial cosmology}
	\label{fig:triangle_ng}
\end{figure}

We have seen in the previous section that $\cng$ and in particular its off-diagonal elements affect significantly the error on cosmological parameters. Because we perform this analysis in a MCMC framework we also have the possibility to check whether the position of the best-fit is affected.

Figure~\ref{fig:triangle_ng} exhibits the 1D and 2D marginalised posterior distributions for $\kmax=0.25\ \invMpc$ in the three covariance cases. As expected from the previous section, the diagonal non-Gaussian contribution has almost no effect on the constraints and the posteriors obtained in that case and with the Gaussian covariance are similar and well centered on the fiducial cosmology. However when adding the off-diagonal terms the posterior drifts away from the fiducial cosmology in particular for $h$ and $m_\nu$ with a shift of the order of $1\sigma$. Note that here the most shifted parameters are $h$ and $m_\nu$, which we remind to be the most and the least constrained parameters respectively.

\begin{figure*}
    \centering
    \includegraphics[scale=0.38]{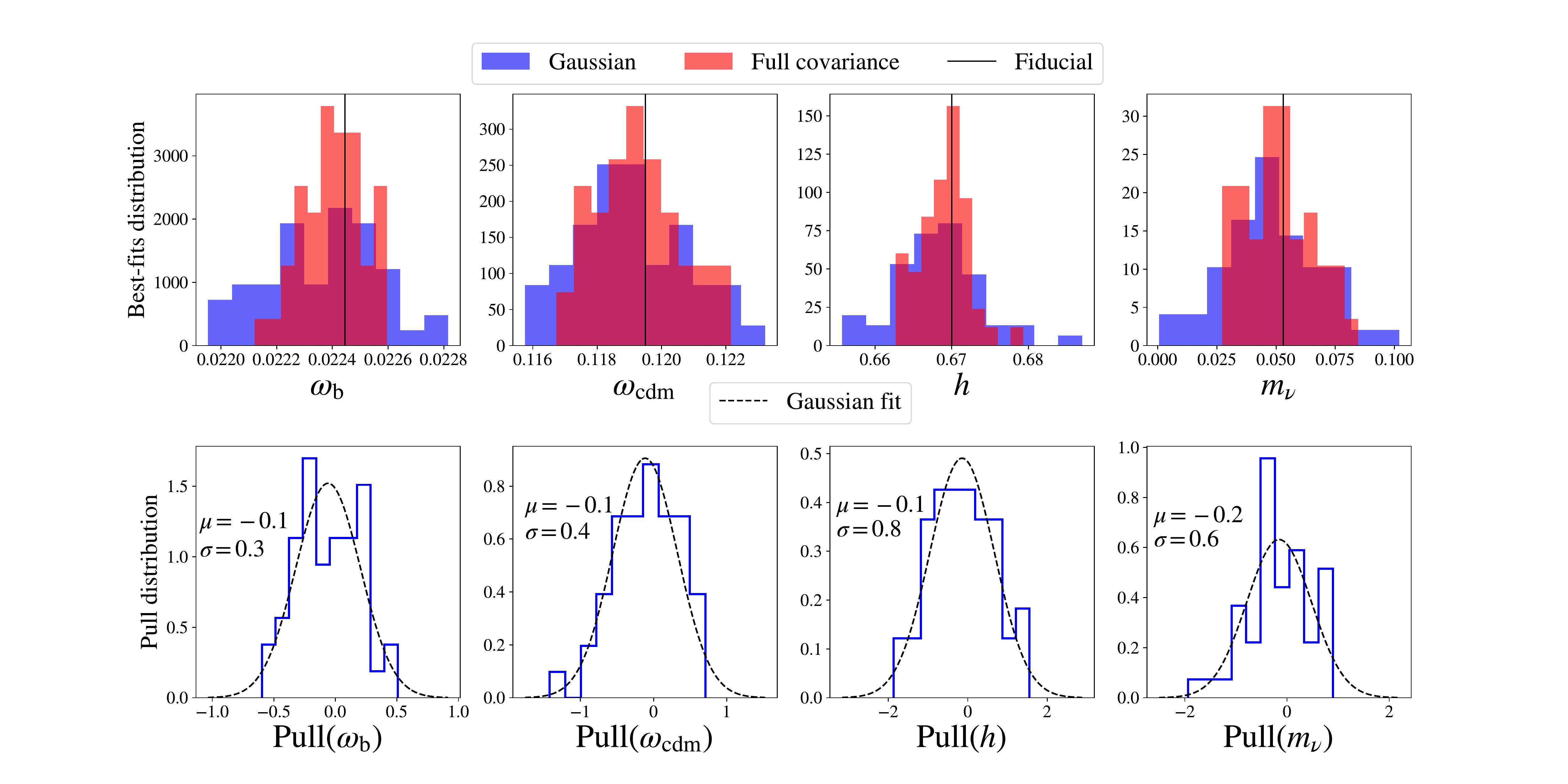}
    \caption{Results from fitting 50 different realisations of the \covhf\ $P(k)$ at $\kmax=0.25\ \invMpc$. {\it Top}: Normalised distribution of best-fits in the case of the Gaussian (blue) and full covariance (red). {\it Bottom}: Normalised distribution of the pull as defined in Eq.~\eqref{eq:pull}. The black dashed line is a Gaussian fit to the histogram. The mean and standard deviation obtained from the fit for each parameter are written in each panel.}
    \label{fig:histo_bf_res}
\end{figure*}

To asses whether this shift of the best-fit is systematic or stochastic we perform the same fit but for 50 different realisations of the data-vector taken from the \covhf\ data-set. In Figure~\ref{fig:histo_bf_res} we show the distribution of the best-fits obtained from these 50 fits using either the Gaussian or full non-Gaussian covariance, as well as the residual between the two distributions normalised by the quadratic sum of the errors obtained in each of the fits. This latter quantity is also known as the pull and defined as 
\begin{equation}\label{eq:pull}
    \mathrm{Pull}(\theta) \equiv \dfrac{(\theta_\mathrm{full}-\theta_\mathrm{Gauss})}{\sqrt{\sigma^2(\theta_\mathrm{full})+\sigma^2(\theta_\mathrm{Gauss})}}.
\end{equation}

First, looking at the distribution of the best-fits we can see that on average the fiducial cosmology is well retrieved for the two covariances. In addition, the distribution of the pull is close to a Gaussian distribution with a mean equal to 0, meaning that the shift of the best-fit is purely stochastic. Performing a Gaussian fit of this distribution we indeed find the fitted mean to be compatible with zero for all parameters. This also gives us the standard-deviation of the pull distribution which can indicates which are the most affected parameters by this stochastic shift and by how much in term of a fraction of their error-bar. As observed above $h$ and $m_\nu$ are the parameters with the higher standard-deviation, $\sigma=0.8$ and 0.6 respectively. We also did the same study for $\kmax=0.2\ \invMpc$ and found the same effect but with lower strength. For that $\kmax$ the standard deviations found from fitting a Gaussian to the pull distribution are at maximum $\sigma=0.5$ and 0.4, still for $h$ and $m_\nu$ respectively.

\citet{sellentin_19} demonstrated that assuming an incorrect covariance matrix could bias parameter inference. Here we showed that this bias exists but does not have a preferred direction over multiple realisations of the data. Still, with real data we only have one realisation of the Universe, the best-fit will differ for one covariance matrix or another and we should trust the value obtained with the correct one. Though as we have seen, it is not easy to asses the accuracy of a covariance matrix.

A last observation we can make about these results is that surprisingly the best-fits are more dispersed in the Gaussian covariance case (c.f. upper panel of Figure~\ref{fig:histo_bf_res}) despite the fact that the error on the parameters were found to be smaller. Intuitively we could think that the dispersion of the best-fits obtained from a given number of realisations should match the error estimated from the width of the posterior, at least on average. 

But, actually it should not be the case if the likelihood assumed to perform the parameter inference does not match the underlying true one,  for example because the assumed covariance matrix is not the correct one. Indeed this is demonstrated in appendix B of \citet{payerne_22}, where they show that, for a Gaussian likelihood, the ensemble covariance matrix of the estimated parameters ({\it i.e.} obtained from an ensemble of estimated best-fit values) and the individual one ({\it i.e.} obtained from the width of an individual posterior) are different if the assumed data covariance matrix deviates from the true one. More precisely, if the true data covariance matrix is larger than the assumed one then the ensemble variance of parameters is larger than the individual one. This directly applies in our case by identifying the Gaussian covariance matrix to the assumed one and the \covmos\ covariance matrix to the true one and we exactly observe the consequence predicted by \citet{payerne_22}.

\subsection{Discussion}

To conclude on the impact of $\cng$ on cosmological parameter constraints, we have seen that the posterior distribution is significantly affected, especially by the off-diagonal contribution to the covariance matrix. The width of the posterior is well underestimated for $\kmax>0.15\ \invMpc$ when wrongly assuming a Gaussian covariance. In addition, for $\kmax>0.2\ \invMpc$, the best-fit values also undergo a stochastic shift that can goes up to 1$\sigma$ if an incorrect covariance is assumed.  It is uneasy to understand why some parameters are more affected than others as the most affected ones are both the most and the least well constrained ($h$ and $m_\nu$ respectively). Still, we note that the presence of a tight prior, on $\omegab$ in our case, largely diminishes these effects. 

We remind that the analysis set-up chosen in this work is quite far from what is actually accessible with galaxy clustering surveys because we are considering here the matter field in real space. However, this is what weak lensing surveys are probing through the statistics of galaxy shapes. Although here we are considering 3D $P(k)$ instead of projected quantities along the line of sight as it is the case for weak lensing and we are not sensitive at all to the systematics characteristic of weak lensing analysis, these results on the matter field can be of direct interest to interpret parameter inference with weak lensing. In addition, our results are consistent with \citet{barreira_18a} who shown that the non-Gaussian contribution to the covariance has a non negligible impact on cosmological constraints obtained with weak lensing, though the SSC contribution seems to be the most important one. 

Note that \citet{wadekar_20} also tested the impact of non-Gaussian contributions to the covariance using an analytical covariance accounting for non-linear galaxy bias and RSD in the case of the full-shape galaxy clustering spectroscopic analysis of BOSS DR12 data. They found non-Gaussian covariance effects to be negligible on cosmological parameters because they are absorbed by nuisance parameters (mainly galaxy bias) through marginalisation. This was also observed in \citet{gouyou_21} for SSC in the case of photometric galaxy clustering. In another work, \citet{yu_2022} used the same data as in \citet{wadekar_20} but a hybrid approach for the covariance matrix, combining analytical prediction and simulations to show that the non-Gaussian covariance could be responsible for a 10-20\% increase in cosmological parameters errors\footnote{Apart from the different treatment of the covariance, the other main methodological difference between \citet{yu_2022} and \citet{wadekar_20} is that in the former work they performed a template fit ({\it i.e.} fixing the shape of the power spectrum and fitting for the AP parameters and the velocity growth factor $f\sigma_8$) while in the latter it was a full-shape fit. }. It is thus unclear how important these additional terms to the covariance are. It would be interesting to pursue these studies that have been done with BOSS data, for stage-IV galaxy survey settings as they should be more sensitive to non-Gaussian covariance by probing smaller and more non-linear scales than stage-III surveys such as BOSS. The present work, which is a first step before adding complexity to the analysis, will certainly be useful to interpret future results.

Finally, we only considered our data to be in periodic boxes. In the more realistic case of a survey window function other terms to the covariance should be considered, not only SSC terms which are non-Gaussian but also additional Gaussian terms accounting for window function effects \citep{li_19}. This is left for future work.

%Still, it is interesting as this is what weak lensing surveys are actually through the correlations between the shapes of galaxies.

%It is not trivial at all to predict how these results could translate in the case of the redshift space galaxy power spectrum, the addition of galaxy bias could potentially increase the contribution from $\cng$ on non-linear scales, while RSD could wash out correlation between different modes due to the stochasticity of galaxy peculiar velocities on small scales.

%This further work should definitely be carried out in the future and there is no doubt that the result presented here will    
\section{Conclusions}
\label{sect:conclu}

In order to be able to get a precise measurement of the still undetected total neutrino mass from the LSS of the Universe, upcoming galaxy surveys will need to probe small scales where the matter field presents non-Gaussian features that appear in the covariance of its power spectrum in the form of the trispectrum. In this article we showed that, to be sure to accurately and precisely detect subtle effects on the LSS such as massive neutrino free streaming, we need a almost perfect understanding of the covariance of our observables. 

There are two complementary ways to model the covariance matrix, through analytical computation, which generally requires the use of perturbative expansions, or by directly estimating it from simulations, which must be numerous enough for the estimate not to be too noisy. In this work we focused on the latter path and studied in section~\ref{sect:samnoise} two techniques to reduce the influence of covariance sampling noise in parameter inference, which can severely bias the posterior distribution of cosmological parameters. 

%This was done performing MCMC fits of the matter power spectrum from simulated data based on the \demcov\ N-Body simulations.

First, to reduce the dispersion of best-fits and errors that affect estimated parameters due to covariance sampling noise, we considered \nercome, a non-linear shrinkage covariance estimator. We found this estimator to significantly reduce these two effects but at the cost of a systematic increase of the errors on the parameters, which is of the same order as the reduction of the aforementioned effects. Thus, while \nercome\ can be useful to get a more accurate position of the best-fit when only a low number of simulations is available, the error-bars it results in cannot be trusted.

We also tested how well Fisher forecast based analytical prescription were performing at predicting sampling noise effects. While the formula from \citet{dodelson_13} is in good agreement with the actual dispersion of the best-fit values of parameters, we found the prediction of the variance on the variance and the bias on the variance by \citet{taylor_13} and \citet{percival_13} to be higher than what is observed and subdominant with respect to the former effect. At the same time we checked if having non-Gaussian posterior distributions would change these results, as it is generally the case when trying to measure the total neutrino mass. We found no major difference with respect to the Gaussian posterior case.

%, because its posterior distribution is cut by the physical lower bound $M_\nu > 0$

The second technique we considered to deal with a low number of simulations to estimate the covariance is \covmos. We found the best-fit values of cosmological parameters estimated using a \covmos\ covariance matrix to be unbiased, thus demonstrating the potential of this method for the analysis of future galaxy surveys. However we could not precisely evaluate how accurate parameters error bars are when using \covmos. This is something that should be quantitatively evaluate in a future work.

%, a novel Monte-Carlo approach able to generate very large samples of realisations of a non-Gaussian field with a given power spectrum and 1-point PDF.

%Future galaxy surveys will go deep into non-linear regime, thus will be sensitive to non-Gaussian cov. This is mandatory to measure neutrino mass. To be sure to detect such small effect as neutrinos need for a perfect treatment of the covariance. One way is to estimate cov from simu. But noisy.

%In this paper we studied how to deal with noise + the impact of non-Gaussian cov through MCMC of real space matter P(k) estimated from DEMNUni simu including neutrinos. 

%\begin{itemize}
    %\item We studied \nercome\ and saw that in the end, the reduction of best-fit dispersion does not compensate for the systematic increase of the errors. Thus nice for some studies and to have an idea of the importance of bf disp, but not reliable to have accurate errors.
    %\item We tested analytical predictions for sampling noise effects and found that \citet{dodelson_13} is ok but not \citet{taylor_13}. \citet{percival_13} is subdominant. We also tested in the non-Gaussian posterior limit and no major issue were found
    %\item We tested \covmos\ in a parameter estimation for the first time. It works great. We could not test how well cosmological errors are recovered with \covmos.
%\end{itemize}

Then, in section~\ref{sect:ng_cov} we focused on the impact of the non-Gaussian term of the power spectrum covariance, $\cng$ (c.f. Eq.~\eqref{eq:pkcov}), coming from the LSS non-linear evolution, on cosmological parameters posterior distribution. We shown that the diagonal part of $\cng$ does not affect cosmological constraints but it is rather the correlation between the different modes ({\it i.e.} the off-diagonal part) that has an impact. When considering this non-Gaussian covariance the error-bars on cosmological parameters increase significantly for $\kmax>0.15\ \invMpc$, when compared to the Gaussian covariance case. In particular, the error increase for $h$ and $m_\nu$ goes up to 20\% and 30\% respectively for $\kmax=0.25\ \invMpc$. This happens when we restrict $\omegab$ to a BBN prior, otherwise the increase is even higher (30\% and 40\% respectively). 

Furthermore, we saw that $\cng$ could also affect the position of the best-fit for $\kmax>0.2\ \invMpc$. Indeed, by performing MCMC's on multiple realisations of the data, we found that when assuming the incorrect covariance the best-fit values of parameters undergo a stochastic shift with respect to the value obtained with the true covariance. At  $\kmax=0.25\ \invMpc$ and for the most affected parameters, $h$ and $m_\nu$, this shift is of the same order as their statistical errors. It is thus important to account for non-Gaussian covariance not only to get the true errors on cosmological parameters but also to get an accurate estimate of their central value.  

As we considered the matter power spectrum in real space in a periodic box, the non-Gaussian term $\cng$ here refers to the trispectrum contribution correlating Fourier modes inside the observed volume (c.f. Eq.~\eqref{eq:trispectrum}). Thus, SSC as well as contribution from a survey window function in the Gaussian term was not accounted for. Also, shot-noise related terms \citep{lacasa_17HM} were neglected as we were working with a negligible shot-noise level. 

% With a good cov like covmos, we studied the effect of non-Gaussian cov on parameters errors. Here only matter in real space with no window, so only two terms: $\cg$ and $\cng$. No window in $\cg$ and no SSC. Also no shot-noise related terms. We found that 
% \begin{itemize}
%     \item It is the off-diagonal terms that have an impact
%     \item Large impact on errors for $\kmax>0.15\ \invMpc$
%     \item Impact on bf for $\kmax>0.2\ \invMpc$. Stochastic shift of the bf with a std dev of the order of the error for $\kmax=0.24\ \invMpc$.
%     \item $h$ and $m_\nu$ are the most impacted params while they are the least and most well constrained
%     \item Tight prior on $\omega_b$ prevents from any of these effects.
% \end{itemize}

The covariance matrix is a complex and rich statistical object and its accurate prediction is one of the most important challenge that upcoming galaxy surveys will have to face. There are many different possible approaches (analytical, semi-analytical, N-Body simulations, approximated simulations...) to model a variety of terms arising from non-linear clustering, galaxy bias, shot-noise, survey window function, observational systematics and more. It is therefore mandatory for the data analysis of future surveys such as {\it Euclid}, DESI or LSST, to know which terms are dominant and must be accounted for. To understand all this hierarchy of covariance terms it is interesting to tackle the problem step by step, by adding complexity as we go further. This is what the present work aims and it should certainly be pursued in the future.    

%Covariance is complex and one of the most important challenge for future surveys, because many possible approaches (analytic, simu, hybrid) to deal with a lot of different terms (Gaussian, window, mom-Gaussian, SSC, shot-noise). It can be useful to take the problem step by step to better understand as we add complexity. This is what this works aims, but it should clearly be pursued to be closer to the analysis set-up of future surveys.

\begin{acknowledgements}
The authors would like to thank Constantin Payerne and Calum Murray for useful discussions. PB and SGB were supported by CNES, focused on Euclid mission. This work received support from the French government under the France 2030 investment plan, as part of the Excellence Initiative of Aix-Marseille University -A*MIDEX (AMX-19-IET-008 -IPhU). The DEMNUni simulations were carried out in the framework of ``The Dark Energy and Massive-Neutrino Universe" project, using the Tier-0 IBM BG/Q Fermi machine and the Tier-0 Intel OmniPath Cluster Marconi-A1 of the Centro Interuniversitario del Nord-Est per il Calcolo Elettronico (CINECA). We acknowledge a generous CPU and storage allocation by the Italian Super-Computing Resource Allocation (ISCRA) as well as from the coordination of the ``Accordo Quadro MoU per lo svolgimento di attività congiunta di ricerca Nuove frontiere in Astrofisica: HPC e Data Exploration di nuova generazione'', together with storage from INFN-CNAF and INAF-IA2.
\end{acknowledgements}


%%%%%%%%%%%%%%%%%%%%%%%%%%%%%%%%%%%%%%%%%%%%%%%%%%%%%%%%%%%%%%%%%%%%
\bibliographystyle{aa}
\bibliography{bibliography}
%%%%%%%%%%%%%%%%%%%%%%%%%%%%%%%%%%%%%%%%%%%%%%%%%%%%%%%%%%%%%%%%%%%%

\appendix
\section{Impact of the isotropic Alcock-Paczyński effect on parameter degeneracy}\label{app:mpch}

In this appendix we show with an example the difference between performing a fit of the full-shape of the $P(k)$ with $\invMpc$ or $\mr{Mpc}^{-1}$ units. As explained in section~\ref{subsect:obs_space} this can be understood as accounting or not for the isotropic Alcock-Paczyński effect.

For this example we ran a MCMC fit of the $P(k)$, at the 5 redshifts, of one realisation of the \covhf\ data-set and used 100~000 realisations to estimate the covariance matrix. Note that we apply the scale cut in the $\invMpc$ space in all cases such that the number of $k$ bins considered is the same and the scale ranges probed are equivalent in both observable spaces.

\begin{figure}[h!]
	\centering
	\includegraphics[scale=0.46]{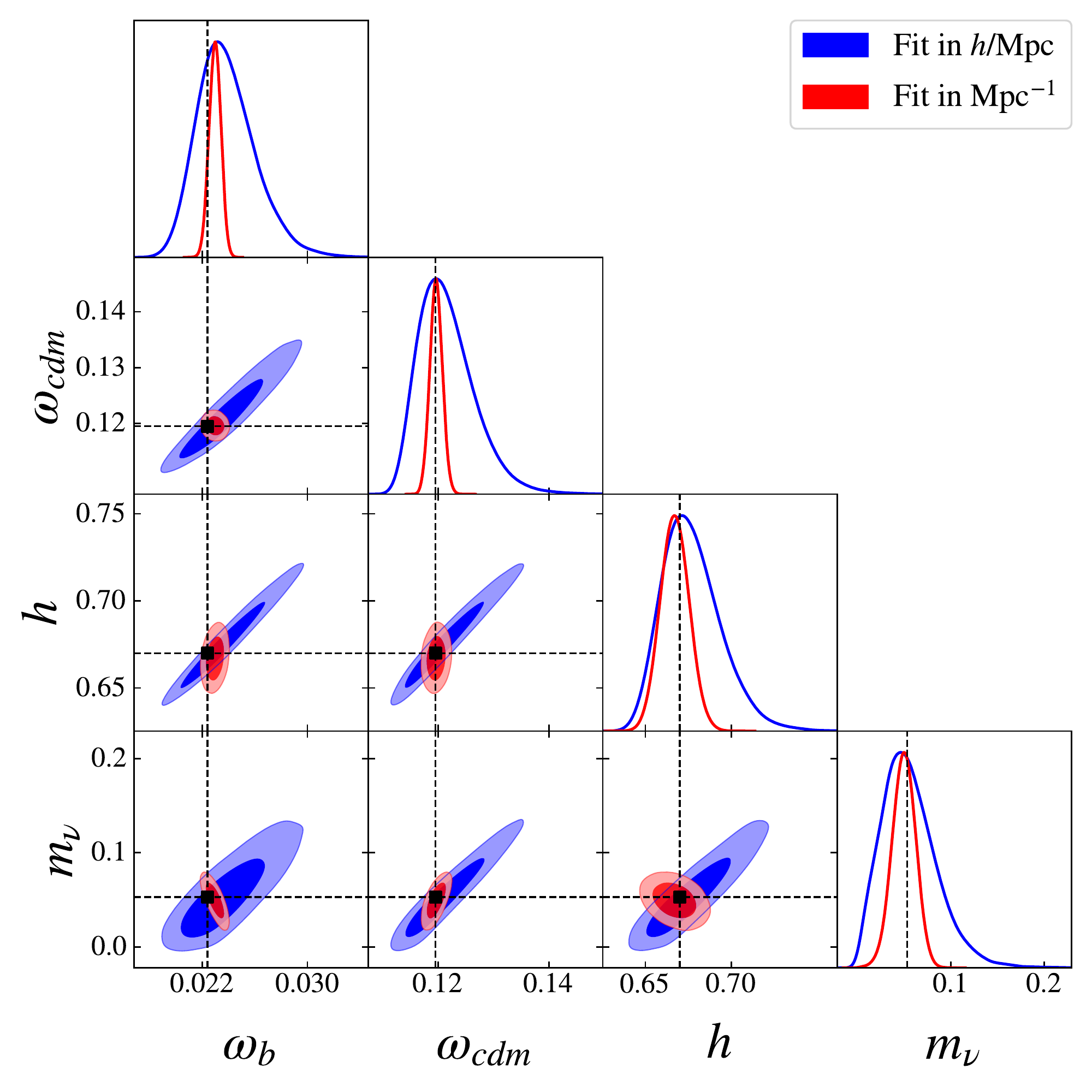}
    \caption{2D and 1D marginalised posteriors, obtained performing the fit in the $\invMpc$ or $\mr{Mpc}^{-1}$ space. For the 2D posteriors, the 68.3\% and 95.5\% confidence regions are shown. This is the constraints for $\kmax=0.19\ \invMpc$. The black square and dashed lines show the fiducial 16$\nu$ cosmology.}
	\label{fig:triangle_mpch}
\end{figure}

Figure~\ref{fig:triangle_mpch} shows the 2D and 1D marginalised posteriors obtained when performing the fit in $\invMpc$ or $\mr{Mpc}^{-1}$. We can see that releasing the dependence in $h$ in the units of the $P(k)$ and the $k$ grid brings strong degeneracies between all fitted cosmological parameters. Due to this degeneracies the posterior distribution presents non-Gaussian features, especially for $m_\nu$. Here the fit is done using a covariance matrix with negligible sampling noise. However, as we want to study what happens with a noisy covariance matrix, the posterior will surely move around the parameter space due to this noise and thus could present even stronger non-Gaussian features. To have a better control on the shape of the posterior we choose to perform the fits in the $\mr{Mpc}^{-1}$ space.

These strong degeneracies could be broken by applying more restrictive priors on some of the parameters, for example on $\omegab$ with a prior coming from BBN results. In section \ref{sect:samnoise} we don't want the results to be affected by prior effects, but we consider this option in section~\ref{sect:ng_cov}.

%One important reason for that is the fact that 

\section{Non-Gaussian posterior}\label{app:ng_post_0nu}
\begin{figure}[h!]
	\centering
	\includegraphics[scale=0.46]{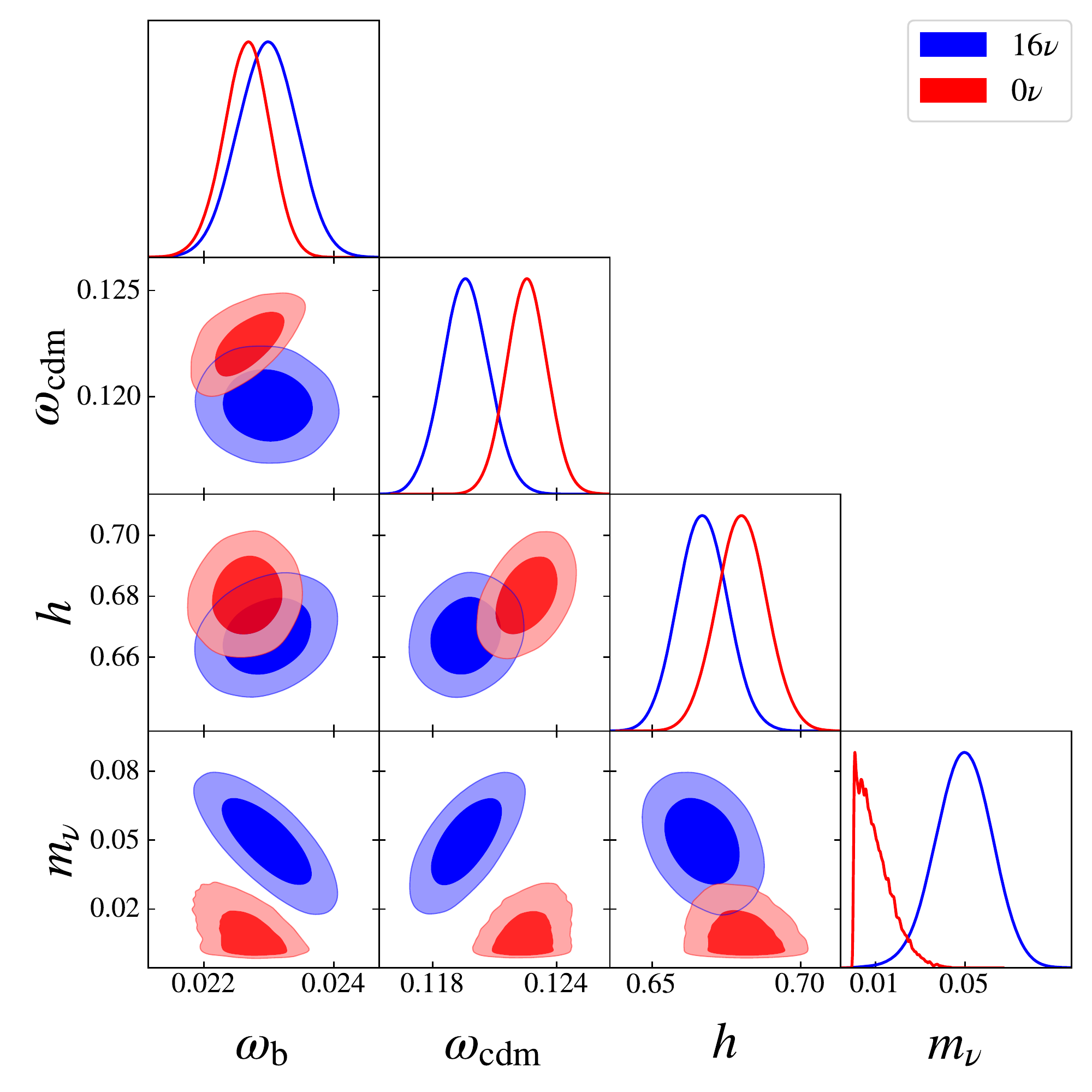}
    \caption{2D and 1D marginalised posteriors, obtained in the 0$\nu$ or the 16$\nu$ cosmology. For the 2D posteriors, the 68.3\% and 95.5\% confidence regions are shown. This is the constraints for $\kmax=0.2\ \invMpc$.}
	\label{fig:triangle_0nu_16nu}
\end{figure}

In this appendix we show how performing a fit in the 0$\nu$ cosmology results in non-Gaussian posteriors.

For this example we ran a MCMC fit of the $P(k)$, at the 5 redshifts, of one realisation of the \covhf\ data-set and used 100~000 realisations to estimate the covariance matrix, for each of the two cosmologies. 

Figure~\ref{fig:triangle_0nu_16nu} shows the 2D and 1D marginalised posteriors we obtain. We can see that as $m_\nu=0$ in the 0$\nu$ cosmology, the MCMC is forced to sample the parameter space which is close to the physical prior $m_\nu>0$. Thus, the marginalised posterior of $m_\nu$ is highly non-Gaussian. Because of the correlation between $m_\nu$ and $\omegab$ the latter also exhibits a non-Gaussian marginalised posterior distribution.

\end{document}